\documentclass[aps, prd, superscriptaddress, nofootinbib, preprintnumbers]{revtex4}

\usepackage{ulem} 
\usepackage{cancel}
\usepackage{graphicx,color}

\usepackage{amsmath}
\usepackage{amssymb}
\usepackage{autobreak}
\usepackage{caption}
\usepackage{enumerate}
\usepackage{svg}
\usepackage{diagbox}
\graphicspath{{./}{figures/}}
\usepackage{amssymb}
\usepackage{ulem} 

\usepackage{epstopdf}
\usepackage{subfigure}
\usepackage{cancel}

\usepackage[breaklinks=true]{hyperref}

\begin{document}

\title{

Probing an intrinsically flavorful ALP 
via 
tau-lepton flavor physics

}

\author{Chuan-Xin Cui}\thanks{{\tt cuicx1618@mails.jlu.edu.cn}}
\affiliation{Center for Theoretical Physics and College of Physics, Jilin University, Changchun, 130012,
China}

\author{Hiroyuki Ishida}\thanks{{\tt ishidah@pu-toyama.ac.jp}}
	\affiliation{Center for Liberal Arts and Sciences, Toyama Prefectural University, Toyama 939-0398, Japan}

\author{Shinya Matsuzaki}\thanks{{\tt synya@jlu.edu.cn}}
\affiliation{Center for Theoretical Physics and College of Physics, Jilin University, Changchun, 130012,
China}

\author{Yoshihiro Shigekami}\thanks{{\tt shigekami@sjtu.edu.cn}}
\affiliation{Tsung-Dao Lee Institute and School of Physics and Astronomy,
Shanghai Jiao Tong University, 800 Dongchuan Road, Shanghai, 200240 China}

\begin{abstract}

Any axionlike particle (ALP) intrinsically possesses flavorful couplings 
to the standard model (SM) fermions 
arising 
as a consequence of the right-handed flavor rotation within the SM. 
In this paper we discuss this intrinsically flavored ALP, and explore 
the correlation of a minimal set of the couplings in a view of coherence in flavor physics observables.
We focus particularly on the tau-lepton flavor violation (LFV).  
The ALP is assumed to be 
tau-philic on a current-eigenstate basis, a la Pecci-Quinn, and is allowed 
to couple also 
to muon and electron only in a  
right-handed specific manner. 
Several LFV processes are generated 
including radiative tau decays 
and also anomalous magnetic moments of electron and muon. 
We first pay attention to two separated limits: electron scenario with the ALP coupled to 
tau which mixes only with right-handed electron, and muon scenario as the muonic counterpart of the electron scenario. 
It turns out that those scenarios are highly constrained by existing experimental 
limits from the LFV processes and $(g-2)$s, to require 
a mu or electron - tau flipped feature in 
the mass eigenbasis when coupled to the ALP. 
We then examine a hybrid scenario combining the two separated scenarios, 
and find a fully viable parameter space on 
the ALP mass-photon coupling plane, which limits the ALP mass to be around $(1.7 - 10)$ GeV 
and the ALP decay constant $f_a$ to be (12.8 -- 67.9) GeV. 
Discrimination of the present ALP from other light LFV particles are also discussed.  
We find that the same-sign multilepton signal at Belle II is 
a smoking-gun to probe the present ALP signal, and
the polarization asymmetry in LFV radiative $\tau$ decay 
is a punchline, which definitely predicts preference of  
the right-handed polarization, in sharp contrast to the prediction of the SM plus massive Dirac neutrinos having 
the highly left-handed preference, and also other light-new physics candidates 
with the same mass scale as the present ALP. 
Possible model-building to underlie the intrinsically-flavorful third-generation specific ALP 
is also briefly addressed. 


\end{abstract}

\maketitle


\section{Introduction}

It is well-known that axion, emergent as a consequence of the Pecci-Quinn (PQ) symmetry breaking, 
gives a solution for the strong CP problem~\cite{Peccei:1977hh,Peccei:1977ur,Weinberg:1977ma,Wilczek:1977pj}, which is one of important issues left in the standard model (SM). 
The axion generically leaves phenomenological footprints in experiments and astrophysical 
observations (see Refs.~\cite{Irastorza:2018dyq,DiLuzio:2020wdo} as a recent review), which arise associated with 
presence of the coupling to $U(1)_A$ anomaly, including not only 
the gluonic, but also the electromagnetic form, 
and also to the SM-fermion (axial) currents charged under the PQ symmetry. 
Axions having such phenomenological probes are collectively called 
axionlike particles (ALPs), and have been keeping the position as 
one of attractive new physics (NP) candidates, 
no matter which  the ALP may or may not solve the strong CP problem.

Of interest is to notice that 
since the ALP carries an axial charge which would share with the SM fermions, 
any ALP possesses intrinsic flavorful couplings to fermions, which arise from   
the right-handed flavor rotation redundant in the framework of the SM. 
This is the ``minimal flavor violation" (MFV) for the ALP~\footnote{
This ``MFV" is outside the standard definition of the MFV, 
because the flavorful ALP couplings to fermions also break 
the family symmetry $U(3)^5$, in addition to the Yukawa sector in the SM. 
}.  
Plenty of studies on the flavor-physics probes of the ALP have so far been performed. 
To our best knowledge, however, 
all those studies have been worked with 
assuming additional flavorful 
couplings beyond the ``MFV" framework above: ALP couplings to SM fermions are 
given as a mixture of the intrinsically 
flavored part and some underlying ultraviolet (UV) contribution beyond the SM. 
If one considers some UV physics of the ALP which would not give any 
flavorful structure, it is still unsure how the intrinsically flavored ALP, constrained by the ``MFV" scheme, could survive or leave a smoking-gun in the flavor physics 
today, or in the future. 
This is worth exploring, and 
is our central motivation in the present 
work.

In the ballpark of the flavorful-ALP research, the ALP mass range around MeV--GeV 
has been focused particularly on, 
because 
it can be covered 
by the prospected upcoming experiments, say, the Belle II~\cite{Belle-II:2020jti},
and 
can also be desired to explain the current deviation from the SM in anomalous magnetic 
moment of muon, $(g-2)_\mu$~\cite{Muong-2:2021ojo,Muong-2:2006rrc,Aoyama:2020ynm}~\footnote{The SM prediction to $(g-2)_\mu$ has so far extensively been studied. See, e.g.,  Refs.~\cite{Aoyama:2012wk,Aoyama:2019ryr,Czarnecki:2002nt,Gnendiger:2013pva,Davier:2017zfy,Keshavarzi:2018mgv,Colangelo:2018mtw,Hoferichter:2019mqg,Davier:2019can,Keshavarzi:2019abf,Kurz:2014wya,Melnikov:2003xd,Masjuan:2017tvw,Colangelo:2017fiz,Hoferichter:2018kwz,Gerardin:2019vio,Bijnens:2019ghy,Colangelo:2019uex,Blum:2019ugy,Colangelo:2014qya,Davier:2010nc}.}, allowing 
the $a$-$\mu$ coupling~\cite{Bauer:2019gfk,Ge:2021cjz,Keung:2021rps} and/or the $a$-photon coupling producing the significant 
Barr-Zee (BZ) type contribution~\cite{Barr:1990vd}.  
It is interesting to also note that this ``sweet" mass range coincides with  
loopholes, yet unconstrained rooms, on the ALP mass-photon coupling space
(See, e.g., Refs.~\cite{Jaeckel:2015jla,Bauer:2017ris,Dolan:2017osp,Beacham:2019nyx,Dobrich:2019dxc}). 
Thus, the ALP with the mass ranged around MeV--GeV has been motivated well to be a 
one promising-NP particle to be probed at the Belle II.

As argued in the recent literature~\cite{Buen-Abad:2021fwq}, however, 
it might not be plausible 
for the ALP to have flavorful couplings which allow the $\mu-e$ conversion~\cite{SINDRUMII:1993gxf,SINDRUMII:1996fti,SINDRUMII:2006dvw},  
particularly because of the severe constraint from the 
muonium-antimuonium oscillation~\cite{Endo:2020mev}. 
One way out allowing the ALP to be flavorful may be to apply folklore to the ALP.  
That is like ``{\it The third-generation is special, and treated differently from the first two.}~\cite{Frampton:1994rt,Frampton:1995wf}", which has often been motivated in modeling with a flavor symmetry. 
Since the ALP would be associated with $U(1)$ axial symmetry (or PQ symmetry) of fermions, hence has the type of mass couplings to fermions, 
the fermion-coupling strengths should generically be aligned to the fermion mass hierarchy, 
on the basis where the PQ symmetry or the PQ current is defined. 
Along this criterion, it would be most plausible to assume that 
the ALP predominantly couples to the third-generation fermions, because they are most massive.


Note also that experimental constraints on lepton flavor violation (LFV) processes involving tau lepton 
are milder than those for muon decay processes, like $\mu \to e \gamma$, 
hence the LFV couplings related with tau lepton is allowed to be larger. 
Tau-LFV processes with ALP have been extensively studied in the  literature~\cite{Davidson:1981zd,Wilczek:1982rv,Berezhiani:1989fp,Heeck:2017xmg,Calibbi:2020jvd,Haghighat:2021djz}. 
Therefore, such a third-generation specific ALP with the ``MFV" can potentially be viable 
in light of searches for tau-LFV processes at the Belle II, within the ``sweet", yet unconstrained room (loopholes) above, with keeping high enough 
sensitivity to $(g-2)_\mu$ as well as $(g-2)_e$~\footnote{ 
There have been several experimental proposals presented to investigate the loopholes~\cite{Alekhin:2015byh,Feng:2018pew,Belle-II:2018jsg,Ishida:2020oxl}. 
All those proposals are based on parametrically flavorful ALP couplings, in contrast to 
the present model highly constrained by the ``MFV" criterion. 
}.  

In this paper, we discuss a third-generation specific and simplified flavorful ALP with the ``MFV",  focusing on the tau-LFV, and explore the intrinsic-flavorful coupling correlations, in light of the Belle II experiment. 
The ALP is assumed to be tau- and bottom-philic on the base where the ``PQ" charge is defined,  and can also couple to muon and electron 
by the ``MFV" arising from the right-handed flavor rotation within the SM.
Thus, the ALP couplings to muon and electron are right-handed specific. 

We first consider two simplified and separated limits: electron scenario with the ALP coupled to tau which mixes only with right-handed electron, and muon scenario as the muon counterpart of the electron scenario. 
It turns out that those scenarios are highly constrained by existing experimental 
limits from the LFV processes and $(g-2)$s, to require 
a mu or electron - tau flipped feature in 
the mass eigenbasis when coupled to the ALP.  

We then examine a hybrid scenario combining the two separated scenarios, 
and find a fully viable parameter space on 
the ALP mass-photon coupling plane, which limits the ALP mass $\sim(1.7 - 10)$ GeV 
and the ALP decay constant $f_a$ to be (12.8 -- 67.9) GeV. 
This ALP can be probed only by measurement 
of $\tau \rightarrow \mu \gamma$ and/or $\tau \rightarrow e \gamma$ at Belle II, 
and cannot be explored by other prospected experiments, including long-lived particle 
searches. 
Discrimination of the present ALP from other light LFV particles are also discussed.  

We find that 
the same-sign multilepton signal at Belle II is a smoking-gun to probe the present ALP, 
and the polarization asymmetry in $\tau \rightarrow \mu \gamma$ and/or $\tau \rightarrow e \gamma$ is a punchline, which gives a definite prediction of   
the right-handed polarization due to the ``MFV", in sharp contrast to the prediction of the SM plus massive Dirac neutrinos having the highly left-handed preference, and also other light-new physics candidates, which 
are promisingly probed at the Belle II.

This paper is organized as follows. 
In Sec.~\ref{sec:model}, we start with a generic setup for the ALP couplings, 
from which the tau-bottom specific ALP with ``MFV" arises, 
and derive the explicit form of the couplings relevant to the LFV processes 
as well as $(g-2)_\mu$ and $(g-2)_e$ discussed in Sec.~\ref{sec:lepFCNC}. 
Sec.~\ref{sec:results}. 
shows constraints and predictions for two separated scenarios: 
the electron scenario and muon scenario. 
In Sec.~\ref{sec:hybrid}, we employ the hybrid scenario and show 
the viable parameter space projected onto the ALP mass-photon coupling plane. 
Then, we propose a smoking-gun and a punchline to discriminate the present ALP from other Belle II-targeted 
NP candidates: 
that is the polarization asymmetry in the radiative tau LFV decay processes. 
Sec.~\ref{sec:summary} is devoted to summary of the present paper, and 
discussion on several issues related to the future prospect along the current work. 
Possible model-building to underlie the present third-generation specific ALP 
is also briefly addressed.

\section{ALP with ``MFV"} 
\label{sec:model}

We begin by considering the ALP couplings to SM quarks and charged leptons. 
We assume the CP invariance for the ALP couplings in a current basis on which a PQ-like charge 
is defined. 
We consider a minimal set of the ALP couplings 
described by the following effective quark and lepton mass-matrix terms:
\begin{equation}
 {\cal L}_{\rm mass} + \mathcal{L}_{a ff} =  
 -\bar{q}_{i}m_q^{ij}q_j-\bar{\ell}_{i}m_{\ell}^{ij}\ell_j-i \frac{a}{f_a}\bar{q}_i \gamma_5 C_q^{ij} q_j -i \frac{a}{f_a}\bar{\ell}_i \gamma_5 C_{\ell}^{ij} \ell_j 
\,,    \label{la1}
\end{equation}
where $q$ = $(u,d,s,\cdot \cdot \cdot, t)^{T}$ and $\ell$ = $(e,\mu,\tau)^{T}$ are the quark and 
lepton fields, 
respectively; $a$ is represented as the ALP field; 
$f_a$ is the decay constant of ALP; the sum over the repeated indices are taken into account.   
We have introduced the ALP-fermion coupling terms as a matrix form, 
$C_{q,\ell}$, which are taken to be real to keep the CP invariance at this point. 
For minimality, we assume the coupling matrices to be diagonal in the flavor basis: 
\begin{align} 
C_{q} &={\rm diag}\{ Q_u m_u, Q_d m_d,\cdots , Q_t m_t \} \, , 
\notag\\ 
C_{\ell} & ={\rm diag} \{ Q_e m_e, Q_{\mu} m_{\mu},Q_{\tau} m_{\tau}\} \,, 
\end{align} 
with the flavor-dependent constants $Q_{q,\ell}$, which 
can be regarded as the PQ-like charges for the corresponding fermions.

Next, we move on to the mass eigenstate basis of fermions. 
The criterion of minimality allows the flavor mixing supplied only from the SM sector, i.e., 
that is the 
intrinsic flavorful source, what we call the ``MFV" for ALP and is dubbed ``MFV". 
Recall the fermion-mass diagonalization in the SM, which is worked out 
by bi-unitary (left and right) transformations of the mass matrices. 
The degree of freedom of the left-handed rotation is completely fixed by 
the Cabibbo-Kobayashi-Maskawa (CKM)~\cite{Cabibbo:1963yz,Kobayashi:1973fv} 
and Pontecorvo-Maki-Nakagawa-Sakata (PMNS) angles~\cite{Pontecorvo:1957qd,Maki:1962mu}, 
while the right-handed one is unfixed within the SM, to be left as 
hidden, redundant and unphysical parameters. 
When the ALP is present and has the chiral couplings to SM fermions, as in Eq.~(\ref{la1}), 
the redundant right-handed rotation actually becomes physical. 
This is only the flavor-violating source for the ALP within the present ``MFV" framework, 
and breaks the parity, allowing the ALP to couple to SM fermions,  
not only in an axial (pseudoscalar) form, but also in a vector (scalar) form.


The rotation matrix of right-handed lepton fields, $U_R^{\ell}$, can be defined in a way similar to the PMNS  matrix. 
We set the CP phase $\delta_{\rm CP}$ to be zero, because the CP violation of lepton sector is not our current concern. 
The rotation matrix $U_R^{\ell}$ is thus parametrized as 
\begin{equation}
    U_R^{\ell}=\left(\begin{array}{ccc}
c_{12} c_{13} & s_{12} c_{13} & s_{13}  \\
-s_{12} c_{23}-c_{12} s_{13} s_{23}  & c_{12} c_{23}-s_{12} s_{13} s_{23}  & c_{13} s_{23} \\
s_{12} s_{23}-c_{12} s_{13} c_{23}  & -c_{12} s_{23}-s_{12} s_{13} c_{23}  & c_{13} c_{23}
\end{array}\right)
\,, \label{UR}
\end{equation}
where $c_{ij} = \cos \theta_{ij}$ and $s_{ij} = \sin \theta_{ij}$. 
By performing this rotation, 
the Lagrangian terms in Eq.~(\ref{la1}) in the mass eigenstate basis  
are cast into the form:
\begin{equation}
    \mathcal{L}^{\prime} \ni - \bar{q}_{i}m_{q_i}\delta_{ij} q_j-\bar{\ell}_{i}m_{\ell_i}\delta_{ij} \ell_j 
    - 
    i\frac{a}{f_a}\bar{q}_i \left( (g_V^q)_{ij}+(g_A^q)_{ij} \gamma_5 \right) q_j 
    - 
    i\frac{a}{f_a}\bar{\ell}_i \left( (g_V^\ell)_{ij}+(g_A^\ell)_{ij} \gamma_5 \right) \ell_j
\,,    \label{mass}
\end{equation}
where the vector and axial couplings are given as 
\begin{align}
    g_V^{\ell}=\frac{C_\ell U_R^\ell - {U_R^\ell}^{\dagger} C_\ell}{2}, \quad  g_A^{\ell}=\frac{C_\ell U_R^\ell + {U_R^\ell}^{\dagger} C_\ell}{2}, \quad
     g_V^q=0, \quad  g_A^q=C_q
     \,. \label{gVA}
\end{align}

The ALP-photon coupling can be induced through quark and charged-lepton loops 
at the one-loop order. 
When the ALP and photons are onshell, the coupling reads \cite{Bauer:2020jbp}
\begin{align} 
{\cal L}_{a \gamma \gamma} 
&= 
C_{\rm \gamma \gamma}^{\rm eff} \frac{\alpha}{4 \pi}\frac{a}{f_a}F_{\mu \nu} \Tilde{F}^{\mu \nu}, 
\end{align} 
where 
\begin{align} 
    \quad C_{\gamma \gamma}^{\rm eff} \equiv 
    \sum_{q_i=u,d,\cdot \cdot \cdot,t} 3 (Q_{q_i}^{\rm em})^2 \frac{(g_A^q)_{q_i q_i}}{m_{q_i}} B_1 \left(\frac{4m_{q_i}^2}{m_a^2} \right)+ \sum_{\ell_i=e, \mu, \tau} \left(Q_{\rm \ell_i}^{\rm em} \right)^2 \frac{(g_A^{\ell})_{\ell_i \ell_i}}{m_{\ell_i}} B_1\left(\frac{4m_{\ell_i}^2}{m_a^2} \right) 
\label{Cgammagamma}
\end{align}
where $F_{\mu\nu}$ is the photon field strength; the dual field strength $\Tilde{F}^{\mu \nu}$ is defined as $\Tilde{F}^{\mu \nu} \equiv \frac{1}{2}\epsilon^{\mu \nu \alpha \beta} F_{\alpha \beta}$, with $\epsilon^{0 1 2 3}$ = $+1$; 
$\alpha$ is the fine-structure constant of electromagnetic coupling; 
$Q^{\rm em}_{q_i(\ell_i)}$ denotes the electric charge (in unit of e) for $i-$ quark ($\ell$ - lepton). 
The loop function $B_1(x)$ is given as 
\begin{equation}
    B_1(x)=-xf^2(x), \quad f(x)= 
    \begin{cases} 
    \displaystyle \sin^{-1}\frac{1}{\sqrt{x}} \hspace{24.3mm} {\rm for} \qquad x\geq 1\\ 
    \displaystyle \frac{\pi}{2}+\frac{i}{2} \ln\frac{1+\sqrt{1-x}}{1-\sqrt{1-x}} \hspace{7mm} {\rm for} \qquad x < 1
    \end{cases}
\,. 
\end{equation}

The loop function $B_1(x)$ applied in Refs.~\cite{Bauer:2020jbp, Cornella:2019uxs} is $1-xf^2(x)$, with an extra constant term "1". 
The difference comes from the model setups~\footnote{
For discussion on this discrepancy, see also Ref.~\cite{Bauer:2017ris}. 
}. 
The present ALP model is assumed to have a minimal set of ALP couplings, 
so that the ALP coupling to photon only arises from one-loop charged fermions, as in Eq.(\ref{Cgammagamma}). 
This setup has been realized by requiring 
a ``bare" coupling $C_{\gamma \gamma}^{\rm bare}$ to cancel 
the electromagnetic $U(1)$ axial anomaly arising due to the anomalous $U(1)_A$ 
rotation~\footnote{
The most general ALP interactions include 
the three types: the mass coupling as in 
Eq.(\ref{la1}), derivative couplings to 
vector and axialvector currents, and 
the ``topological-charge" type ($F\tilde{F}$). One of three can be 
removed by adjusting the axial rotation angle. In this context, the present ALP 
model has assumed to remove the derivative type of couplings by the axial rotation 
in the beginning. 
}. 
If those terms had not been made cancelled each other, by going beyond the present 
minimal setup, 
we would have the same result on $B_1(x)$ as in Refs.~\cite{Bauer:2020jbp, Cornella:2019uxs} 
with the extra term ``1". 
However, as far as the ALP mass around $\sim {\cal O}$(GeV) is concerned, which is the present target mass range, 
the numerical evaluation on amplitudes involving this $C_{\gamma\gamma}^{\rm eff}$ coupling would not substantially be altered whether the term ``1" is included or not. 

It is convenient to further give the relation to the effective ALP-lepton coupling $c_{\ell_i \ell_j}^{\ell}$ 
adopted in Ref.~\cite{Bauer:2020jbp}. 
Since we focus on the tau-bottom specific ALP, 
we set $Q_e=Q_{\mu}=0$ and $Q_{\tau}=1$. 
Given this benchmark, 
the effective coupling $c_{\ell_i \ell_j}^{\ell}$ takes the form
\begin{align}
    \text{flavor diagonal}&: \quad \quad c_{\tau \tau}^{\ell} =\frac{1}{m_{\tau}}(g_A^{\ell})_{\tau \tau},\quad c_{e e}^{\ell}=c_{\mu \mu}^{\ell}=0 
    \,, \notag \\ 
     \text{flavor off-diagonal}&: \quad \quad  c_{\tau \ell_i}^{\ell} =\frac{\sqrt{2}}{m_{\tau}}\sqrt{\mid (g_V^{\ell})_{\tau \ell_i} \mid^2 + \mid (g_A^{\ell})_{\tau \ell_i} \mid^2} \quad (\ell_i \neq \tau),\quad c_{e \mu}=0 
     \,. \label{cl-couplings}
\end{align}

\section{Relevant Leptonic processes}
\label{sec:lepFCNC}

In this section, we list several leptonic processes, including 
lepton-flavor conserving and violating physics, such as  
$(g-2)_{e}$ and $(g-2)_{\mu}$, $\ell_i \rightarrow \ell_j \gamma$ and $\ell_i \rightarrow \ell_j \ell_k \ell_k$. These processes will give constraints on model parameters later, 
in Sec.~\ref{sec:results}.

\subsection{Anomalous magnetic moments: $(g-2)_{\mu}$ and $(g-2)_e$}

\subsubsection{muon $g-2$}

\begin{figure}[t]
\centering
\subfigure[Barr-Zee loop]{
\begin{minipage}[t]{0.25\linewidth}
\centering
\includegraphics[width=0.90\linewidth]{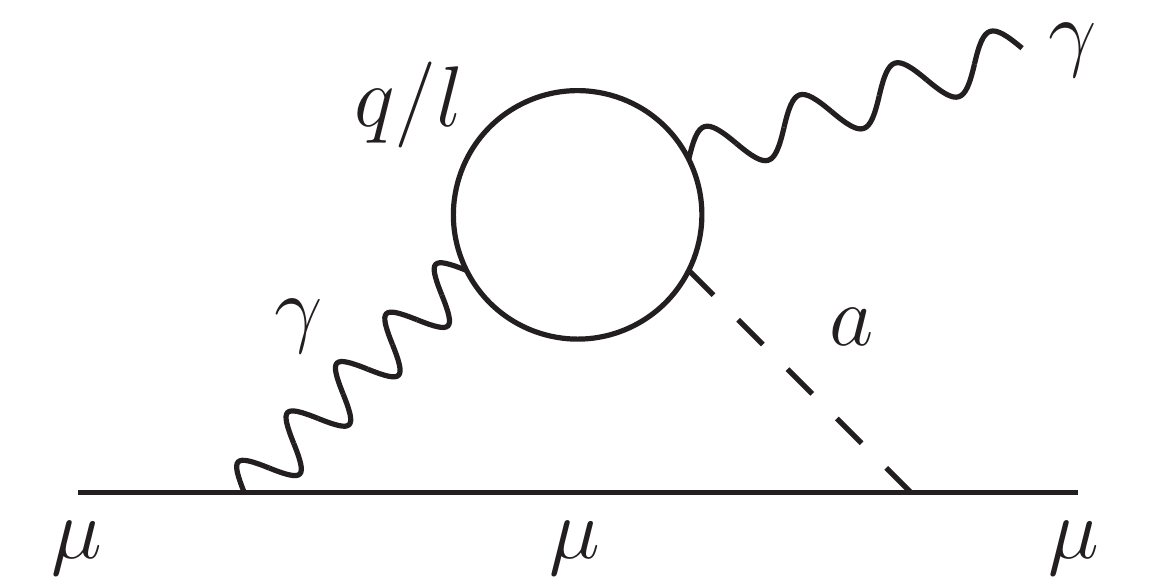}  
\end{minipage}%
}%
\subfigure[arch loop]{
\begin{minipage}[t]{0.25\linewidth}
\centering
\includegraphics[width=0.90\linewidth]{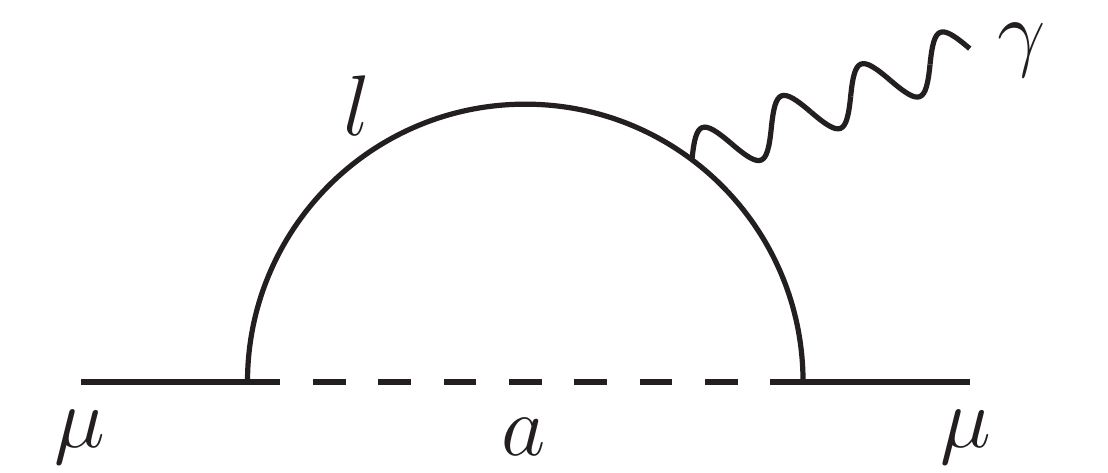}
\end{minipage}%
}%
\centering
\captionsetup{justification=raggedright}
\caption{Illustration of the ALP contributions to $\Delta a_{\mu}$: BZ type (Left) and arch type (right). 
In the BZ type diagram 
the white blob for the $a$-$\gamma$-$\gamma$ vertex is evaluated by the effective coupling $C_{\gamma\gamma}^{\rm eff}$ in Eq.(\ref{Cgammagamma}). }
\label{BZ-arch-graphs}
\end{figure}

The latest measurement on the anomalous magnetic moment of muon, $a_{\mu}\equiv (g-2)_{\mu}/2$, 
reported by Fermi lab~\cite{Muong-2:2021ojo}, tells us 
that the combined average with previous result at Brookhaven~\cite{Muong-2:2006rrc} is 4.2$\sigma$ deviation from the SM prediction $a_{\mu}^{\rm SM}=116 591 810(43)\times 10^{-11}$~\cite{Aoyama:2020ynm} as: 
\begin{equation}
    \Delta a_{\mu} \equiv a_{\mu}^{\rm exp}-  a_{\mu}^{\rm SM} =(2.51 \pm 0.59)\times 10^{-9}
\,. \label{amu-exp}
\end{equation} 
To this discrepancy, the ALP can contribute as the NP term, which 
 can be written as
\begin{equation}
    \Delta a_{\mu}^{\rm NP}= \Delta a_{\mu}^{\rm BZ} + \Delta a_{\mu}^{\rm arch}
    \label{amu_BZarch}
\,, 
\end{equation}
involving two kinds of diagrams: the BZ and arch loop diagrams as depicted in Fig.~\ref{BZ-arch-graphs}. 
The contributions from these two type of diagrams can be evaluated as~\cite{Cornella:2019uxs,Bauer:2017ris,Lindner:2016bgg}
\begin{align}
     \Delta a_{\mu}^{\rm BZ} &=-\frac{m_{\mu}^2}{16 \pi^2 f_{a}^2} \frac{2\alpha}{\pi} c_{\mu \mu}^\ell C_{\gamma \gamma}^{\rm eff} \Big{[}\ln\frac{(4\pi f_a)^2}{m_{\mu}^2}-h_2(x_{\mu})\Big{]} 
     \,, \label{a-mu-BZ} \\ 
     \Delta a_{\mu}^{\rm arch} &= \frac{m_{\mu}^2}{8 \pi^2 f_a^2} \sum_{f = e, \mu, \tau}\Big{(} -(g_V^\ell)_{\mu f} (g_V^\ell)_{f \mu} I_{f,1}^{++}+ (g_A^\ell)_{\mu f} (g_A^\ell)_{f \mu} I_{f,1}^{+-} \Big{)}
\,, 
\end{align}
where $x_{\mu}\equiv m_{a}^2/m_{\mu}^2$ and the loop functions are
\begin{align}
     h_2(x) &:=1+\frac{x^2}{6}\ln x -\frac{x}{3}+\frac{x+2}{3}\times 
     \begin{cases}
    \displaystyle \sqrt{x(4-x)}\cos^{-1}\frac{\sqrt{x}}{2} \quad x<4 \\
    \displaystyle -\sqrt{(x-4)x} \ln \frac{\sqrt{x}+\sqrt{x-4}}{2} \quad x>4
    \end{cases} \,, \\
   I_{f, \, 1}^{(\pm)_1 (\pm)_2} &:= I_{f, \, 1} \left[ m_{\ell_i}, (\pm)_1 m_{\ell_j}, (\pm)_1 m_{\ell_f}, m_a \right] \nonumber \\
   &= 
   \displaystyle \int \! dx dy dz \, \delta(1 - x - y - z) \frac{x \left( y + (\pm)_1 z \frac{m_{\ell_j}}{m_{\ell_i}} \right) + (\pm)_2 (1 - x) \frac{m_{\ell_f}}{m_{\ell_i}}}{- x y m_{\ell_i}^2 - x z m_{\ell_j}^2 + x m_a^2 + (1 - x) m_{\ell_f}^2} \, ,
    \label{g3}
\end{align}
where $m_{\ell_i}$ and $m_{\ell_j}$ correspond to charged-lepton masses in initial and final states for the process, respectively, and $m_{\ell_f}$ is the charged-lepton mass in the loop.

In evaluating the BZ diagram in Eq.~(\ref{a-mu-BZ}), we have introduced a cutoff scale and replaced it simply by $4 \pi f_a$. 
Presence of the logarithmic divergence is due to the current simplified 
evaluation of the diagram, in that  the $a$-$\gamma$-$\gamma$ vertex function is replaced by  
momentum-independent onshell-effective coupling $C_{\gamma\gamma}^{\rm eff}$, 
in such a way that the BZ contribution can be evaluated as an effective one-loop graph. 
If the BZ diagram were computed exactly at the two-loop level, we would have a finite result as in the literature~\cite{Chang:2000ii,Buttazzo:2020vfs}. 
No matter which evaluation is applied, 
however, the BZ contribution vanishes in the current tau-specific model  
with ALP being flavorful via only the right-handed mixing angles, which leads to $c_{\mu\mu}^l = 0$ from Eq.(\ref{cl-couplings}): $\Delta a_\mu^{\rm BZ}=0$.

\subsubsection{electron $g-2$}

A very recent accurate measurement on the fine-structure constant of the electromagnetic coupling has 
reported 
the latest value of the SM prediction to the anomalous magnetic moment of electron, $a_e^{\rm SM}=1159652180.252(95)\times 10^{-12}$~\cite{Morel:2020dww}. Comparing to the direct experiment measurement~\cite{Hanneke:2008tm}, we see about 
1.6 $\sigma$ discrepancy: 
\begin{equation}
    \Delta a_{e} \equiv a_{e}^{\rm exp}-  a_{e}^{\rm SM}  =(4.8 \pm 3.0)\times 10^{-13} 
    \label{pae}
\,. 
\end{equation} 
It is worth noting that the two deviations, $\Delta a_{e}$ and $\Delta a_{\mu}$, have the same signs. 
Actually, the latest measurement \cite{Morel:2020dww} fairly disagrees with 
the previous results without any clear reason. 
Therefore, here we also quote the previous value of $\Delta a_{e}$~\cite{Garcia-Martin:2008xjr,Hanneke:2010au}, 
the sign of which is opposite compared to $\Delta a_{\mu}$:
\begin{equation}
    \Delta a_{e}|_{\rm old} =(-8.7 \pm 3.6)\times 10^{-13}
    \label{poae}
\,. 
\end{equation} 
Similarly to $\Delta a_{\mu}$ in Eq.~(\ref{amu_BZarch}), 
the ALP contributes to $\Delta a_{e}$ as the NP term, which is split into two terms: 
\begin{equation}
    \Delta a_{e}^{\rm NP} = \Delta a_{e}^{\rm BZ} + \Delta a_{e}^{\rm arch}
\,, \label{Delta-a-e}
\end{equation}
where 
\begin{equation}
     \Delta a_{e}^{\rm BZ} =-\frac{m_{e}^2}{16 \pi^2 f_{a}^2} \frac{2\alpha}{\pi} c_{e e}^\ell C_{\gamma \gamma}^{\rm eff} \Big{[} \ln\frac{(4\pi f_a)^2}{m_{e}^2}-h_2(x_{e})\Big{]}
 \,, \label{BZ-ae}
    \end{equation} 
 \begin{equation}
     \Delta a_{e}^{\rm arch} = \frac{m_{e}^2}{8 \pi^2 f_a^2} \sum_{f = e, \mu, \tau}\Big{(} -(g_V^\ell)_{e f} (g_V^\ell)_{f e} I_{f,1}^{++}+ (g_A^\ell)_{e f} (g_A^\ell)_{f e} I_{f,1}^{+-} \Big{)}
\,. \label{arch-ae}
 \end{equation}
Again, the present BZ contribution vanishes because $c_{ee}^l=0$, from Eq.(\ref{cl-couplings}): $\Delta a_e^{\rm BZ} =0$.

\subsection{Radiative LFV: $\ell_i \rightarrow \ell_j \gamma$}

The branching ratio for $\ell_i \rightarrow \ell_j \gamma$ decays process can be evaluated as \cite{Bauer:2019gfk, Buttazzo:2020vfs} 
\begin{align}
  Br(\ell_i \rightarrow \ell_j \gamma)
  &=\frac{\Gamma (\ell_i \rightarrow \ell_j \gamma)}{\Gamma _{\ell_i}} 
  \,, \notag\\ 
  \Gamma (\ell_i \rightarrow \ell_j \gamma)
  &=\frac{\alpha m_{\ell_i}^5(c_{\ell_i \ell_j}^\ell)^2}{4096 \pi^4 f_a^4} \left|c_{\ell_i \ell_i }^\ell g_1(x_{\ell_i})+\frac{2\alpha}{\pi}\sum_{f_i=u,d,..., \mu, \tau}\frac{(g_A^f)_{f_i f_i}}{m_{f_i}}f\left(\frac{m_a^2}{m_{\ell_i}^2},\frac{m_a^2}{m_{f_i}^2}\right)
\right|^2
 \,, 
  \label{g1}
\end{align}
where $x_{\ell_i} \equiv m_a^2/m_{\ell_i}^2$ and
$\Gamma_{\ell_i}$ is the total width of lepton $\ell_i$. 
The loop functions showing up in Eq.~(\ref{g1}) are:
\begin{align}
    g_1(x) &= 2\sqrt{4-x}x^{\frac{3}{2}} \cos^{-1} \frac{\sqrt{x}}{2}+1-2x+\frac{3-x}{1-x}x^2 \ln x, \\
    f(u,v) &=\int_0^1 dx dy dz \frac{ux}{u \bar{x}+uvxyz\bar{z}+vz\bar{z}x^2\bar{y}^2}
    \label{g2}
\end{align}
where $\bar{x}(\bar{y},\bar{z})$ is the shorthand notation for $x(y,z)-1$. 
Note that in the present ALP model, there are only bottom and $\tau$ contributions to the second term in Eq.~\eqref{g1}. See Eqs.~\eqref{Brtuy_e} and \eqref{Brtuy_mu}. In contrast to $\Delta a_\mu^{\rm BZ}$ and $\Delta a_e^{\rm BZ}$ in Eqs.(\ref{a-mu-BZ}) and (\ref{BZ-ae}), 
the overall coupling for the BZ graph contributions is set by 
the flavor-off diagonal one, $c_{l_i l_j}^l$, which leads to nonzero contributions (via Eq.(\ref{cl-couplings}))
to the radiative LFV processes. However, it will turn out that 
the BZ terms is actually required to be highly suppressed by 
phenomenological arguments.

Here we have kept the leading term in expansion with respect to large $m_{\ell_i}$, so 
that only the $\ell_i$ lepton contribution has been taken into account in evaluating 
the arch loop, because other terms will either involve flavor changing couplings twice, 
or be suppressed by the smaller lepton masses, hence are all likely to be subdominant.

\subsection{Purely leptonic LFV: $\ell_i \rightarrow \ell_j \ell_k \ell_k$}

Other important and stringent constraints would come from purely leptonic LFV processes, like  $\ell_i \rightarrow \ell_j \ell_k \ell_k$. 
The mass of ALP is fairly sensitive to those processes:   
when $m_a<2m_{\ell_k}$ or $m_a>m_{\ell_i}-m_{\ell_j}$, the ALP can only be produced 
off mass shell, while for $2m_{\ell_k}<m_a<m_{\ell_i}-m_{\ell_j}$, the ALP can be produced at on-shell. For simplicity, we only consider the tree-level contribution from the ALP. 
The explicit expressions of the branching ratios for those processes are given as follows \cite{Cornella:2019uxs}: 
for $ m_a>m_{\ell_i}-m_{\ell_j}$, we have 
\begin{equation}
     \Gamma(\ell_i \rightarrow \ell_j a^{*}  \rightarrow  \ell_j \ell_k \ell_k) =
     \frac{c_{\ell_k \ell_k}^2 c_{\ell_i \ell_j}^2 m_{\ell_i}^3 m_{\ell_k}^2}{256 \pi^3 f_a^4} \varphi_0^{j k}(x_{\ell_i}) 
\,, 
\end{equation}
where 
\begin{align}
    \varphi_0^{jj}(x)&=-\frac{11}{4}+4x- \Big{[} \frac{x^2}{2} \ln \frac{2x-1}{x}-1+5x-4x^2 \Big{]} \ln \frac{x-1}{x}+\frac{x^2}{2}\Big{[} {\rm Li}_2 \left(\frac{x-1}{2x-1} \right)-{\rm Li}_2\left(\frac{x}{2x-1} \right) \Big{]} \,, \\
     \varphi_0^{j\neq k}(x)&=(3x^2-4x+1) \ln \frac{x-1}{x}+3x-\frac{5}{2}
     ; 
\end{align} 
for $ 2m_{\ell_k}<m_a<m_{\ell_i}-m_{\ell_j}$, we have 
\begin{equation}
    \Gamma(\ell_i \rightarrow \ell_j a  \rightarrow  \ell_j \ell_k \ell_k)\approx \Gamma(\ell_i \rightarrow \ell_j a) Br(a \rightarrow \ell_k \ell_k)= \tau_a  \frac{c_{\ell_k \ell_k}^2 c_{\ell_i \ell_j}^2 m_{\ell_i}^3 m_{\ell_k}^2}{256 \pi^2 f_a^4} \left( 1-\frac{m_a^2}{m_{\ell_i}^2} \right)^2 \sqrt{m_a^2-4 m_{\ell_k}^2}
\,, 
\end{equation}
where $\tau_a$ denotes the lifetime of ALP.


\subsection{LFV production of on-shell ALP: $l_i \to l_j a$}

 At tree level, the on-shell ALP can be generated through the decay process $\ell_i \rightarrow \ell_j a$ and its partial width is computed as \cite{Cornella:2019uxs} 
\begin{align}
    \Gamma(\ell_i \rightarrow \ell_j a)=\frac{m_{\ell_i}^3}{32\pi} \left( 1-\frac{m_a^2}{m_{\ell_i}^2} \right)^2\frac{c_{\ell_i \ell_j}^2}{f_a^2}
\,. \label{l-la}
\end{align} 


\subsection{Experimental limits}

In Table \ref{Explim}, we give a summary of the currently available experimental bounds and 
future prospected limits on the tau-LFV processes.

 \begin{table}[!htbp]
 \centering
 \renewcommand{\arraystretch}{1.8}
 \setlength{\tabcolsep}{7mm}
\begin{tabular}{|c|c|c|}
\hline
LFV process & Experiment limit with $90\%$ C.L. & Future prospect  \\ \hline
$\tau \rightarrow e \gamma$ & $3.3\times 10^{-8}$ $\quad$ \cite{Hayasaka:2010np}& $3\times10^{-9}$ $\quad$ \cite{Belle-II:2018jsg}\\ \hline

$\tau \rightarrow \mu \gamma$ & $4.4\times 10^{-8}$  $\quad$ \cite{Hayasaka:2010np}& $10^{-9}$ $\quad$  $\quad$ \cite{MEGII:2018kmf} \\ \hline

$\tau \rightarrow 3 e$ & $2.7\times 10^{-8}$  $\quad$ \cite{Hayasaka:2010np}& $5\times 10^{-10}$ $\quad$ \cite{Belle-II:2018jsg} \\ \hline

$\tau \rightarrow 3 \mu$ & $2.1\times 10^{-8}$  $\quad$  \cite{Hayasaka:2010np}& $4\times 10^{-10}$ $\quad$ \cite{Belle-II:2018jsg}\\ \hline

$\tau \rightarrow e \mu^{+} \mu^{-}$ & $2.7\times 10^{-8}$  $\quad$ \cite{Hayasaka:2010np} & $6\times 10^{-10}$  $\quad$ \cite{Belle-II:2018jsg}\\ \hline

$\tau \rightarrow \mu e^{+} e^{-}$ & $1.8\times 10^{-8}$  $\quad$ \cite{Hayasaka:2010np} & $3\times 10^{-10}$ $\quad$ \cite{Belle-II:2018jsg}  \\ \hline

$\tau \rightarrow e + {\rm inv} $ & $\approx 2.7 \times 10^{-3}$  $\quad$ \cite{ARGUS:1995bjh}&  $-$  \\ \hline

$\tau \rightarrow \mu + {\rm inv} $ & $\approx 5 \times 10^{-3}$  $\quad$ \cite{ARGUS:1995bjh} &  $-$ \\ \hline
\end{tabular}
\captionsetup{justification=raggedright}
\caption{Current and future prospected limits on the relevant tau-LFV processes.} 
\label{Explim}
\end{table}

\section{Two scenarios for bottom-tau specific model}
\label{sec:results}

In this section, we investigate two separated scenarios for the third-generation specific model relevant to the tau-lepton flavor physics, as well as $\Delta a_e$ and/or $\Delta a_\mu$.  We employ the two separated scenarios: the electron scenario and muon scenario. 
For the electron scenario, we turn on only $\theta_{13}$, which connects the tau and electron flavor physics. While for the muon scenario, only $\theta_{23}$ is taken to be nonzero,  to allow tau and muon flavor physics to be correlated. 
We maximize the ALP coupling to bottom quark, by simply assuming 
no flavor mixing with the ALP in the quark sector, i.e., $\theta_q =0$. 
The setup for the lepton sector in these two scenarios 
is summarised in Table~\ref{benTab}. 

\begin{table}[!htbp]
 \renewcommand{\arraystretch}{1.8}
 \setlength{\tabcolsep}{6mm}
\begin{tabular}{|c|c|c|}
\hline
\multicolumn{3}{|c|}{third-generation specific model with $Q_{\tau}=Q_{b}=1$, other $Q_f$=0}                 \\ \hline
scenario          & parameter setup & nonzero leptonic processes \\ \hline
electron scenario & $\theta_{13} \neq 0$,   $\theta_{23}=\theta_{12}=0$        & $\Delta  a_e$, $\tau \rightarrow e \gamma$, $\tau \rightarrow e + {\rm inv}$ 
\\ \hline
muon scenario     & $\theta_{23} \neq 0$, $\theta_{13}=\theta_{12}=0$         & $\Delta a_{\mu}$, $\tau \rightarrow \mu \gamma$, $\tau \rightarrow \mu + {\rm inv}$                        \\ \hline
\end{tabular}
\captionsetup{justification=raggedright}
\caption{The electron and muon scenarios with the corresponding parameter setup in the lepton sector, and the associated LFV processes as well as the lepton $(g-2)$s. }
\label{benTab}
\end{table}

\subsection{Electron scenario}\label{E-scenario}
In this scenario, there are only two nonzero couplings to SM fermions left: 
\begin{align} 
c_{\tau \tau}=\cos{\theta_{13}} \,,  
\qquad 
c_{e \tau}=\sin{\theta_{13}}
\, . 
\end{align}
Then the ALP-photon coupling in Eq.~(\ref{Cgammagamma}) is now simplified to 
\begin{equation}
    C_{\gamma \gamma}^{\rm eff}=\frac{1}{3} B_1 \left(\frac{4m_{b}^2}{m_a^2} \right)+\cos \theta_{13} B_1\left(\frac{4m_{\tau}^2}{m_a^2} \right)
    \label{cyye}
\,. 
\end{equation}
Note that $C_{\gamma \gamma}^{\rm eff}$ does not simply scale with $\theta_{13}$ because it involves the bottom-quark loop contribution, hence the sign of $\cos\theta_{13}$ becomes 
relevant, and $\theta_{13}$ is allowed to take the values in the first and the second quadrants, i.e., 
$\theta_{13} \in [0,\pi]$.

Similarly, we can simplify  
$\Delta a_{e}^{\rm NP}$ in Eq.~(\ref{Delta-a-e}) with Eqs.~(\ref{BZ-ae}) and (\ref{arch-ae}), 
$\tau \rightarrow e \gamma$ in Eq.~(\ref{g1}) with Eq.~(\ref{g2}), and 
$\tau \rightarrow e a$ in Eq.~(\ref{l-la}), to get 
\begin{align}
    \Delta a_{e}^{\rm NP} &=\frac{\sin^2 \theta_{13}}{f_a^2} F_1(m_a) \label{Dae}
    \,, \\
    Br[\tau \rightarrow e a] &=\frac{\sin^2 \theta_{13}}{f_a^2} F_2(m_a) 
    \label{Brtua_e}
    \,, \\ 
     Br[\tau \rightarrow e \gamma] &=\frac{\sin^2 \theta_{13} }{f_a^4} F_3(m_a) \left| \cos \theta_{13} g_1 \left(\frac{m_a^2}{m_{\tau}^2} \right) 
     +\frac{2\alpha}{\pi} \left[ f\left(\frac{m_a^2}{m_{\tau}^2},\frac{m_a^2}{m_{b}^2}\right)+\cos \theta_{13}f\left(\frac{m_a^2}{m_{\tau}^2},\frac{m_a^2}{m_{\tau}^2}\right)\right]
    \right|^2 \label{Brtuy_e} 
\,,  
\end{align} 
where 
$F_1(m_a)$,  $F_2(m_a)$ and $F_3(m_a)$ are defined as 
\begin{align} 
    F_1(m_a) &=\frac{m_{e}^2m_{\tau}^2}{32 \pi^2}\big{(} I_{f, 1}^{++} +I_{f,1}^{+-} \big{)}  \,, \label{F_1} \\
    F_2(m_a) &=\tau_{\tau}\frac{m_{\tau}^5}{32 \pi} \left(1-\frac{m_a^2}{m_{\tau}^2} \right)^{2} \frac{1}{m_{\tau}^2} 
    \,, \\
     F_3(m_a) &=\tau_{\tau}\frac{\alpha m_{\tau}^5}{4096 \pi^4} 
     \,. 
\end{align}
It is interesting to note that the present ALP generically gives a positive contribution to $\Delta a_e$, because $F_1(m_a)$ is positive definite.

\begin{figure}[htbp]
\subfigure[$\Delta a_{e}$]{
\begin{minipage}[t]{0.497\linewidth}
\includegraphics[width=1\linewidth]{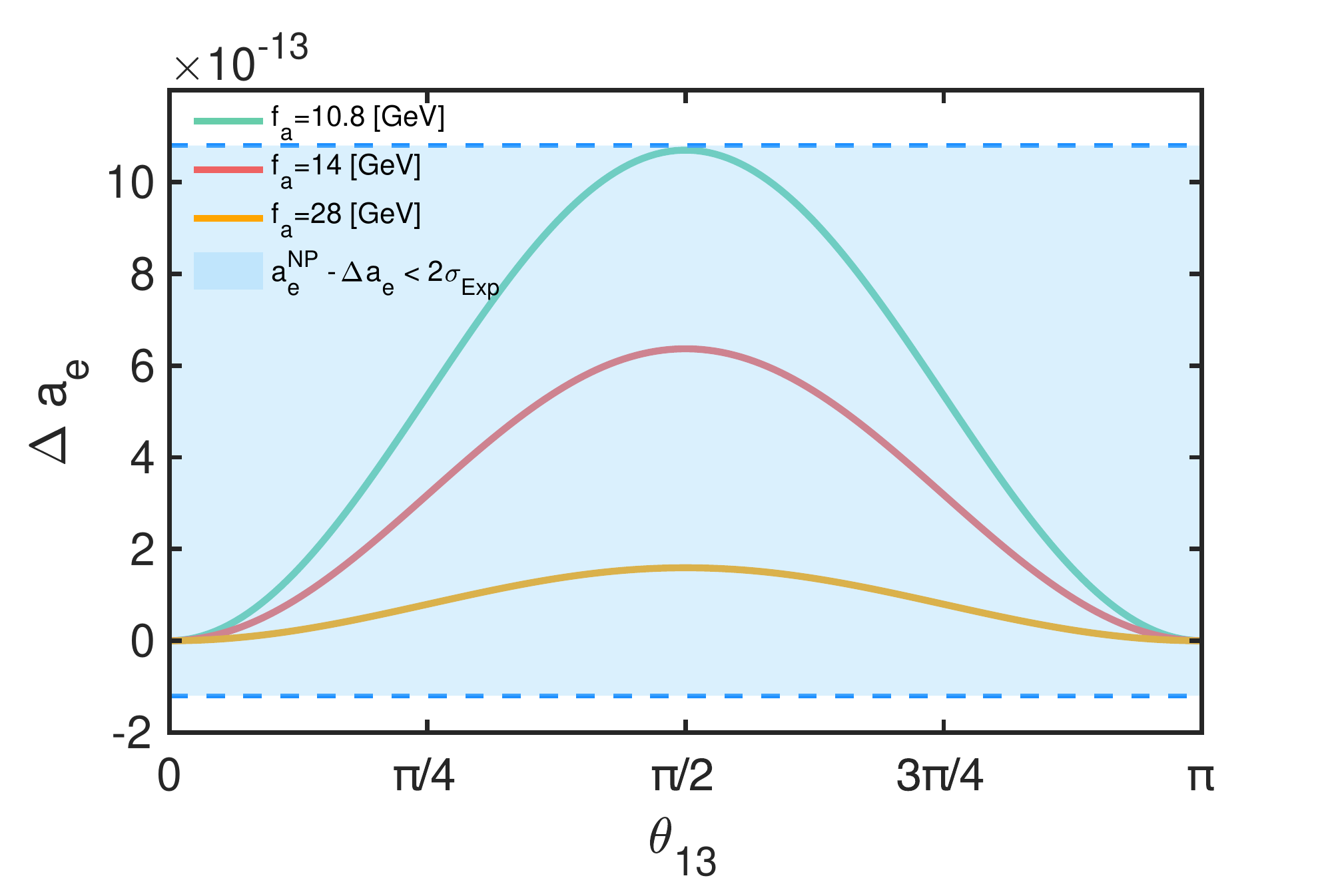}  
\end{minipage}%
}%
\subfigure[$\tau \rightarrow e \gamma$]{
\begin{minipage}[t]{0.49\linewidth}
\includegraphics[width=1\linewidth]{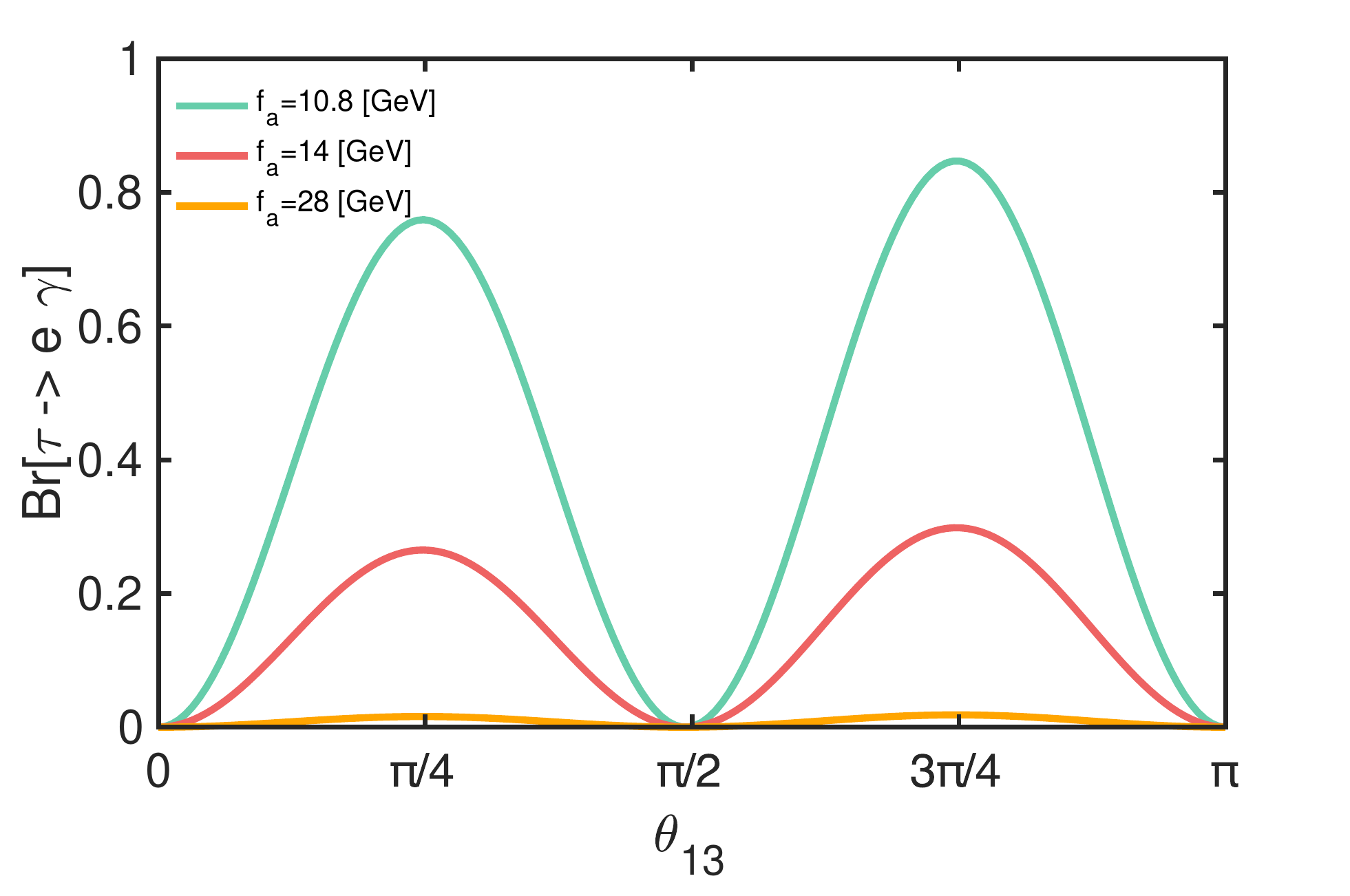}
\end{minipage}%
}%
\captionsetup{justification=raggedright}
\caption{
(Electron scenario): 
$\theta_{13}$ dependence of $\Delta a_{e}$ (left panel) and $Br[\tau \rightarrow e \gamma]$ (right panel). In the left panel, the blue band stands for the currently allowed region 
within the $2\sigma$ deviation read off from Eq.(\ref{pae}).  
The ALP mass $m_a$ has been set to 2 GeV in the plots. } 
\label{ae_ma1}
\end{figure}

In Fig.~\ref{ae_ma1} we show the numerical result on $\Delta a_e$ (left panel) 
and $Br[\tau \rightarrow e \gamma]$ (right panel), 
for $m_a$ = 2 GeV, with $f_a$ varied. 
As seen from the figure, 
$\tau \rightarrow e \gamma$ process tends to be predicted to be too large  ($\mathcal{O}(1)$) to survive the current experiment limit ($\mathcal{O}(10^{-8})$ 
as in Table~\ref{Explim}). There exists a destructive cancellation between 
the arch and BZ diagrams, where the dominant contribution comes from the $\tau$ arch loop. 
Thus $Br[\tau \rightarrow e \gamma]$ approximately scales with $\sin^2\theta_{13} \cos^2\theta_{13}$. 
Hence, to realize small enough $Br[\tau \rightarrow e \gamma]$ while still keeping sizable deviation of $\Delta a_e$, 
$\theta_{13}$ should be fine-tuned to around $\pi/2$, close to 
the exact cancellation between the $\tau$ arch loop and the BZ contributions.  Below we give the precise value of the allowed region for $\theta_{13}$ when $m_a=2$ GeV:

\begin{align} 
    &f_a=10.8 \, \text{GeV},\quad \theta_{13}\simeq 1.55132 - 1.55152, \quad \text{Br}[\tau \rightarrow e \gamma] = (0 - 3.3)\times 10^{-8} \\
    &f_a=14 \, \text{GeV} ,\quad \theta_{13}\simeq 1.54966 - 1.55001, \quad \text{Br}[\tau \rightarrow e \gamma] = (0 - 3.3)\times 10^{-8} \\
    &f_a=28 \, \text{GeV} ,\quad \theta_{13}\simeq 1.54482- 1.54622, \quad \text{Br}[\tau \rightarrow e \gamma] = (0 - 3.3)\times 10^{-8} 
\end{align}

As evident from Eq.~(\ref{F_1}), 
the ALP contribution to $\Delta a_e$ is necessarily positive. 
For $\Delta a_e$ to stay in the region within the $2\sigma$ deviation, 
the decay constant $f_a$ is thus constrained to be larger than 10.8 GeV when $m_a=2$ GeV. 
If we adopt the previous value of $\Delta a_e$, 
$\Delta a_e|_{\rm old} = (-8.7 \pm 7.2)\times 10^{-13}$ in Eq.~(\ref{poae}), 
then the present ALP cannot account for $\Delta a_e$, rather, would be ruled out. 


The ALP gets further constraints coming from $\tau \rightarrow e a$, where 
the onshell ALP is generated through the tau decay process. The straightforward numerical calculation shows that $F_2(m_a)/f_a^2$ in Eq.~(\ref{Brtua_e}) 
gives 
too big branching ratio value compared to the experiment constraint except for $\theta_{13}$ taking the value extremely close to 
$0$ or $\pi$. 
However, $\theta_{13}$ $\sim$ $0$ or $\pi$
would yield  $\Delta a_e \ll  10^{-13}$, 
which would still be consistent with 
current measurement of $\Delta a_e$ within the $2\sigma$ error  
in Eq.(\ref{pae}), though 
being trivial in a sense of no LFV. 
Nevertheless, we shall simply exclude the possibility of this case with 
the onshell ALP production through $\tau \rightarrow e a$ 
~\footnote{There is another important limit on tau-LFV decay processes, like $ \tau \rightarrow e + {\rm inv}$. However, in the present ALP model, this process does not give further constraint on the 
model parameters when we consider the $\tau \rightarrow e a$ process, since the overall factor  $F_2(m_a)/f_a^2$ in Eq.~(\ref{Brtua_e}) is always too big to be survived under experiment constraint. 
Therefore, the present model will always be excluded once we allow the onshell ALP production, 
which is irrespective to whether the ALP decays inside, or outside the detector.}, 
such that the present ALP mass 
is constrained as $m_a$ $>$ $ m_{\tau}-m_e$ $\simeq$ 1.78 GeV.

\subsection{Muon scenario}\label{M-scenario}

In the muon scenario with only nonzero $\theta_{23}$, 
we have two nonzero couplings to SM leptons:  
\begin{align} 
c_{\tau \tau}=\cos{\theta_{23}}, 
\, \qquad 
c_{\mu \tau}=\sin{\theta_{23}}.  
\end{align} 
Then all the analytic formulas for the ALP-photon coupling and relevant LFV amplitudes   take essentially the same form as those in the electron scenario, 
just by replacing $\theta_{13}$ with $\theta_{23}$ in Eqs.~(\ref{cyye})-(\ref{Brtuy_e}) 
and substituting $m_e^2$ with $m_{\mu}^2$ in Eq.~(\ref{F_1}): 
\begin{align}
    \Delta a_{\mu}^{\rm NP} &=\frac{\sin^2 \theta_{23}}{f_a^2} F_1(m_a) \label{Damu}
    \,, \\
    Br[\tau \rightarrow \mu a] &=\frac{\sin^2 \theta_{23}}{f_a^2} F_2(m_a) 
    \label{Brtua_mu}
    \,, \\ 
     Br[\tau \rightarrow \mu \gamma] &=\frac{\sin^2 \theta_{23} }{f_a^4} F_3(m_a) \left| \cos \theta_{23} g_1 \left(\frac{m_a^2}{m_{\tau}^2} \right) 
     +\frac{2\alpha}{\pi} \left[ f\left(\frac{m_a^2}{m_{\tau}^2},\frac{m_a^2}{m_{b}^2}\right)+\cos \theta_{23}f\left(\frac{m_a^2}{m_{\tau}^2},\frac{m_a^2}{m_{\tau}^2}\right)\right]
    \right|^2 \label{Brtuy_mu} 
\,.  
\end{align}

\begin{figure}[htbp]
\centering
\subfigure[$\Delta a_{\mu}$]{
\begin{minipage}[t]{0.51\linewidth}
\centering
\includegraphics[width=1\linewidth]{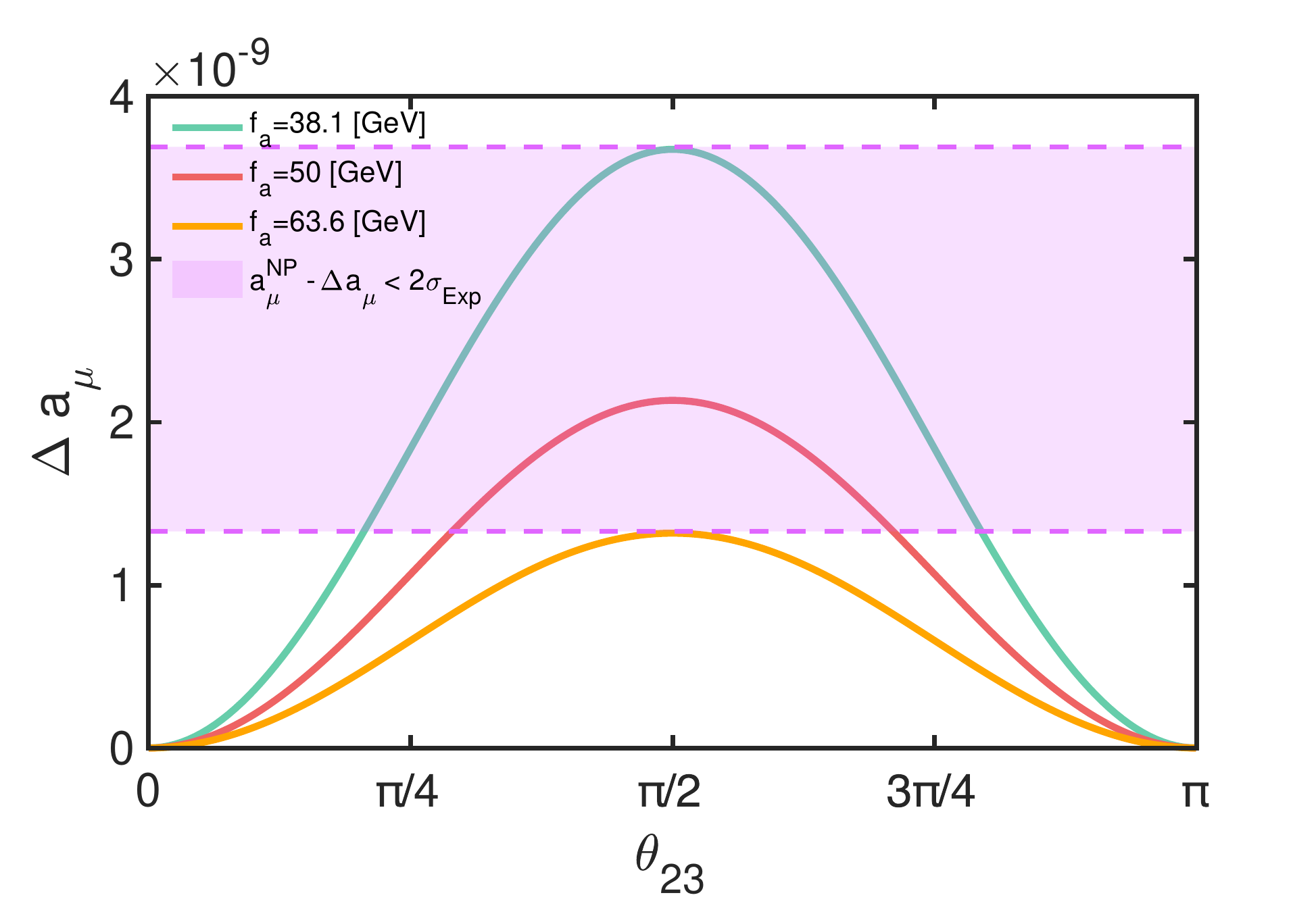}  
\end{minipage}%
}%
\subfigure[$\tau \rightarrow \mu \gamma$]{
\begin{minipage}[t]{0.5\linewidth}
\centering
\includegraphics[width=1\linewidth]{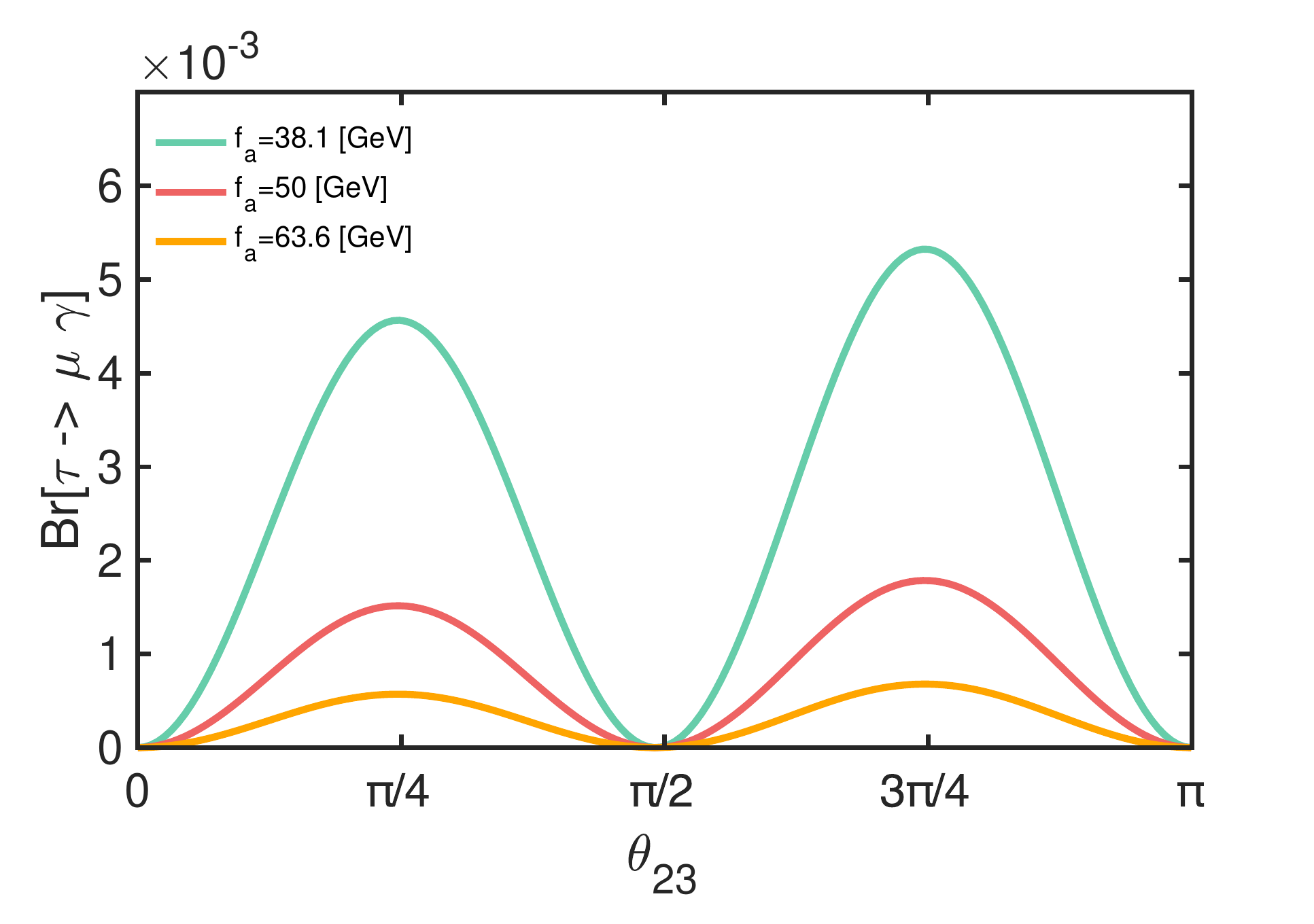}
\end{minipage}%
}%
\centering
\captionsetup{justification=raggedright}
\caption{
(Muon scenario): 
$\theta_{23}$ dependence on $\Delta a_{\mu}$ (left panel) and $Br[\tau \rightarrow \mu \gamma]$ (right panel). 
In the left panel, the purple band stands for the $2\sigma$ allowed range read off from Eq.~(\ref{amu-exp}).  
The ALP mass $m_a$ has been set to 2 GeV in two figures.}
\label{amu_ma1}
\end{figure}

The numerical results on $\Delta a_\mu$ and $Br[\tau \rightarrow \mu \gamma]$ are presented in Fig.~\ref{amu_ma1}, for $m_a = 2$ GeV with $f_a$ varied, 
in a way similar to the electron scenario (Fig.~\ref{ae_ma1}). 
The current $\tau \rightarrow \mu \gamma$ limit only allows tiny window for $\theta_{23}$, where $\theta_{23}$ should be bounded around $\pi/2$ to cause the almost exact cancellation between the $\tau$ arch loop and the BZ term contributions (See Eq.~(\ref{Brtuy_mu})). 
Therefore, with $\theta_{23} \simeq \pi/2$, the decay constant $f_a$ is constrained and 
allowed to vary only in a small range,  38.1 GeV  $-$ 63.6 GeV, 
to yield $\Delta a_\mu$ within the $2\sigma$ deviation (See Eq.~(\ref{Damu})). 
Similarly to the electron scenario, the onshell ALP prodution via $\tau \rightarrow \mu a$ is rule out, and this gives the lower bound on the ALP mass, $m_a \gtrsim 1.67$ GeV. Below we also give the precise $\theta_{23}$ region when $m_a=2$ GeV:

\begin{align} 
    &f_a=38.1 \, \text{GeV},\quad \theta_{23}\simeq 1.54205 - 1.54506, \quad \text{Br}[\tau \rightarrow \mu \gamma] = (0 - 4.4)\times 10^{-8} \\
    &f_a=50 \, \text{GeV} ,\quad \theta_{23}\simeq 1.53923 - 1.54442, \quad \text{Br}[\tau \rightarrow \mu \gamma] = (0 - 4.4)\times 10^{-8} \\
    &f_a=63.6 \, \text{GeV} ,\quad \theta_{23}\simeq 1.53605- 1.54447, \quad \text{Br}[\tau \rightarrow \mu \gamma] = (0 - 4.4)\times 10^{-8} 
\end{align}

Thus, both two scenarios that we have discussed above 
are constrained to have the flavorful ALP coupling with $\theta_{13} \simeq \pi/2$ or $\theta_{23} \simeq \pi/2$.  
It would be intriguing to note that 
this indicates 
the preference of a mu or electron - tau flipped feature in 
the mass eigenbasis when coupled to the ALP. 
In the next section, we will go beyond these specific scenarios, 
and employ a hybrid scenario 
by turning on both $\theta_{13}$ and $\theta_{23}$, and will 
find the surviving parameter space under all the existing relevant constraints. 
Of particular interest is in a hybrid scenario that the currently measured size of deviation in both $\Delta a_e$ and $\Delta a_\mu$ may be explained simultaneously. 

\section{A hybrid scenario}
\label{sec:hybrid}

With both nonzero $\theta_{13}$ and $\theta_{23}$, 
we find the nonzero ALP couplings to SM charged leptons: 
\begin{align} 
c_{\tau \tau} & =\cos \theta_{23} \cos \theta_{13}, 
\notag\\ 
c_{\mu \tau} &=\sin \theta_{23}
\,, \notag\\ 
c_{e \tau} &= \cos \theta_{23} \sin \theta_{13}. 
\label{couplings-hybrid}
\end{align} 
In this section, we focus mainly on the ALP mass within the  
promising Belle II reach with the high sensitivity 
($100\,{\rm MeV}\lesssim m_a \lesssim 10\,{\rm GeV}$), 
and search the surviving ALP-photon coupling space over  
existing and upcoming experimental limits (Part A). 
Then we propose 
a smoking-gun and a punchline  of the ALP, where in particular the latter is the photon polarization 
asymmetry $\lambda_{\gamma}$, clearly distinguishable from other NP 
candidate with the same mass scale, which can also account for 
the current deviations in $\Delta a_e$ and $\Delta a_\mu$ (Part B). 
Finally, we make precise comparison of the ALP with other NP candidates 
in a sense of phenomenology (Part C).

\subsection{Experimental constraints on ALP-photon coupling within Belle II reach}

\begin{figure}[htbp]
\begin{center}
   \includegraphics[width=14cm]{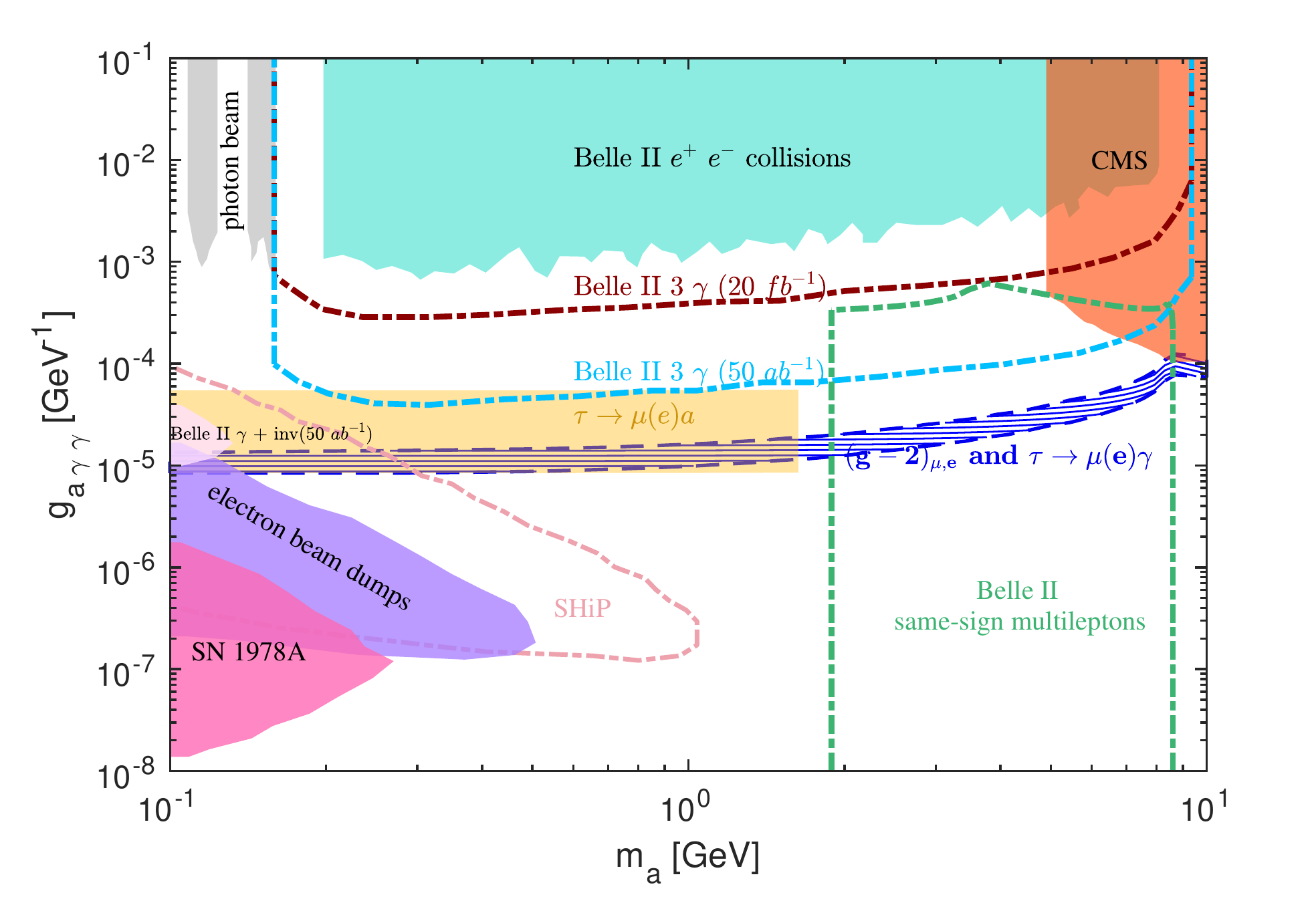}
  \end{center}   
  \captionsetup{justification=raggedright}
\caption{(Hybrid scenario): 
Existing and prospected constraints on ($m_a$, $g_{a \gamma \gamma}$) from various experiments as marked in the plot. 
The allowed parameter space for $(g-2)_{\mu}$ and $(g-2)_e$ 
within the $2\sigma$ error has been displayed 
in the blue slashed area in the center of the figure, 
which also satisfies constraints from $\tau \rightarrow \mu \gamma$ and 
$\tau \rightarrow e \gamma$ processes. The other colored regions 
are excluded from other experiments. 
In particular, the yellow shaded region is excluded by $\Gamma_{\tau}{\rm (LFV)} < \Delta \Gamma_{\tau}$.
Thus, the surviving parameter space corresponds to 
the blue-centered regime labeled as ``$(g-2)_{\mu, e}$ and $\tau \to \mu (e) \gamma$" 
not overlapped with other filled regions with labels of experiments, in color. 
More details on experimental limits and derived ALP phenomenological trends are 
referred to as in the text.} 
\label{parspace}
\end{figure}

In Fig.~\ref{parspace}, we show the exclusion plot for the ALP parameter space spanned by  
$(m_a, g_{a \gamma \gamma})$, where 
\begin{align} 
g_{a \gamma \gamma}\equiv \frac{\alpha |C_{\gamma \gamma}^{\rm eff}|}{\pi f_a}  
\,. \label{g-gamma}
\end{align} 
The existing experimental limits include   
$e^{+} e^{-} \to 3 \gamma$ at Belle II~\cite{Belle-II:2020jti}, 
a photon-beam experiment~\cite{Aloni:2019ruo}, and heavy-ion collisions~\cite{CMS:2018erd};   
electron beam dump experiments \cite{Dolan:2017osp};  
and SN 1987A \cite{Jaeckel:2017tud}. 
We have also incorporated the SHiP prospect \cite{Dobrich:2015jyk}, the future Belle II prospects at 20 ${\rm fb}^{-1}$ and 50 ${\rm ab}^{-1}$ \cite{Dolan:2017osp}, and also the prospected Belle II limit on the same-sign multileptons signal at 50 $ab^{-1}$~\cite{Iguro:2020rby}. 
We find the following features: 

\begin{itemize}
    \item Both $\tau \rightarrow \mu \gamma$ and $\tau \rightarrow e \gamma$ processes require the almost exact cancellation between the $\tau$ arch loop and the BZ contributions, 
    as in the case of the muon and electron scenarios.
    Hence $\theta_{23}$ is severely constrained to be around $\pi/2$, so that the ALP photon coupling $C_{\gamma \gamma}^{\rm eff}$ is dominated by bottom quark loop, $C_{\gamma \gamma}^{\rm eff} \sim \frac{1}{3} B_1(\frac{4m_{b}^2}{m_a^2})$~
    \footnote{Actually, the constraints from both $\tau \rightarrow \mu \gamma$ and $\tau \rightarrow e \gamma$ require $\cos \theta_{13} \cos \theta_{23} \sim 0$. So we have two cases: (i) $\theta_{13}\sim \pi/2$ and $\theta_{23}$ is arbitrary; (ii) $\theta_{23} \sim \pi/2$ and $\theta_{13}$ is arbitrary. In the present analysis, we have adopted the second case, since in the allowed range of $f_a$ (Eq.(\ref{fa_ma})), $\Delta a_{e}$ is necessarily too small 
    irrespective to the size of $\theta_{13}$ (See the discussions below Eq.(\ref{fa_ma})). Therefore, fixing such a potentially free $\theta_{13}$ to a specialized value $\sim \pi/2$, as in the first option above, seems to be unnatural, so we have discarded this case.}. 
    Note that this almost complete cancellation condition is not sensitive to $Br[\tau \rightarrow \mu \gamma]$, nor $Br[\tau \rightarrow e \gamma]$. If $\tau \rightarrow \mu (e) \gamma$ is observed at Belle II with the prospected branching ratio of $\mathcal{O}(10^{-9})$, it can still be consistent with the present ALP. 
    
    \item $f_a$ is constrained and allowed to vary in a narrow range, 
    \begin{align} 
    f_a \simeq (12.7 - 93.3) \, {\rm GeV}, \qquad {\rm for}  \qquad 
    m_a = (0.1 - 10) \, {\rm GeV}.  
    \label{fa_ma}
    \end{align} 
    In such scope of $f_a$, the predicted deviation of $\Delta a_e$ is on the order of magnitude of $\mathcal{O}(10^{-14})$, which is too small that the current experiment limit on $\Delta a_e$ can not give any constraint on $\theta_{13}$.  However, the unconstrained $\theta_{13}$ will not affect the surviving parameter space, because $\theta_{13}$ is always combined with $\cos \theta_{23} \sim 0$ in the relevant 
    coupling form, as clearly seen from Eq.(\ref{couplings-hybrid}). 
    
    \item  The $\tau \rightarrow \mu a$ and $\tau \rightarrow e a$ limits give the lower bound on $m_a$, as in the case of the muon and electron scenarios. 
    $\Gamma(\tau \to \mu a)$ and $ \Gamma(\tau \to e a)$ are always predicted to be too large in the present model, so that the whole on-shell ALP process can also be excluded by a general constraint: $\Gamma_{\tau}(\rm LFV)< \Delta \Gamma_{\tau}$, where $\Gamma_{\tau}(\rm LFV)$ is the partial $\tau$ decay width to which the LFV $\tau$ decay contributes, and $\Delta \Gamma_{\tau}$ is the 1 sigma error in measuring the total $\tau$ decay width, $\Delta \Gamma_\tau \simeq 3.9 \times 10^{-15}$ GeV~\cite{ParticleDataGroup:2020ssz}. The excluded mass range is shown in the yellow region in Fig.~\ref{parspace}. 
    
    \item On the contrary to $\tau \rightarrow \mu a$ and $\tau \rightarrow e a$, 
    the prospected limits from the Belle II same-sign multileptons signals will be a smoking-gun to probe or exclude the present ALP,  which could give the upper bound on $m_a$. The sensitivity is depicted as the green-dashed hatched region in Fig.~\ref{parspace}. The prospected same-sign multileptons signals are quoted from Ref.~\cite{Iguro:2020rby}, where the authors only assume $a-\tau-\mu$ coupling ($c_{\mu \tau}$) without the $a$-photon coupling. In contrast, the present ALP has nonzero ALP-photon coupling, so the constraint on $c_{\mu \tau}$ in Ref.~\cite{Iguro:2020rby} could be milder than what the authors have obtained. Therefore, the green-dashed hatched region in Fig.~\ref{parspace} might merely correspond to the maximal exclusion limit, and the actual allowed space could be wider.  
    Hence, in the high sensitivity region for Belle II, where $100\,{\rm MeV}\lesssim m_a \lesssim 10\,{\rm GeV}$, we find two sets of the allowed ALP mass and decay constant $f_a$, together with 
    the largest allowed range for $\theta_{23}$ in Table~\ref{ma-prediction}. 

\begin{table}[!htbp]
\centering
 \renewcommand{\arraystretch}{1.8}
\begin{tabular}{|c|c|c|c|}
\hline
multilepton constraint & $m_a$(GeV)                & $f_a$(GeV)                  & $\theta_{23}$                              \\ \hline
not included               & $\simeq$ (1.7 $-$ 10)               & $\simeq$ (12.8 $-$ 67.9)              & $\simeq$ (1.42 $-$ 1.55)                    \\ \hline
included                & $\simeq$ (1.67 $-$ 1.88), (8.58 $-$ 10) & $\simeq$ (41.0 $-$ 67.9), (12.8 $-$ 24.3) & $\simeq$ (1.54 $-$ 1.55), (1.42 $-$ 1.47) \\ \hline
\end{tabular}
\,. 
\captionsetup{justification=raggedright}
\caption{The allowed parameter region for $m_a$, $f_a$ and $\theta_{23}$}
\label{ma-prediction}
\end{table}

\end{itemize}

The corresponding predicted values of Br[$\tau \rightarrow \mu \gamma$] and Br[$\tau \to e \gamma$] are shown in Table~\ref{BR-prediction}.

\begin{table}[!htbp]
\centering
 \renewcommand{\arraystretch}{1.8}
\begin{tabular}{|c|c|c|}
\hline
multilepton constraint & Br[$\tau \rightarrow \mu \gamma$]] & Br[$\tau \to e \gamma$]      \\ \hline
not included               & (0 $-$ 4.4)$\times 10^{-8}$                & (0 $-$ 6.7)$\times 10^{-9}$              \\ \hline
included                & (0 $-$ 4.4)$\times 10^{-8}$               & (0 $-$ 1.5)$\times 10^{-9}$, (0 $-$ 6.7)$\times 10^{-9}$ \\ \hline
\end{tabular}
\,.
\captionsetup{justification=raggedright}
\caption{Predicted branching ratio value for $\tau \to \mu (e) \gamma$ processes.}
\label{BR-prediction}
\end{table}

Thus, 
the present ALP in the hybrid scenario can be probed by $\tau \to \mu \gamma$ and/or $\tau \to e \gamma$, but cannot have sensitivity to $e^+ e^- \to 3 \gamma$, 
even with higher statistics $(50\, {\rm ab}^{-1})$, as seen from Fig.~\ref{parspace}. 
It is also remarkable to notice that the surviving parameter space points to  
a part of the currently unexplored ``loopholes", which cannot be explored even by 
the upcoming long-lived particle search experiments, such as SHiP, 
and only the $\tau \to \mu(e) \gamma$ can probe. 
In the next subsection, we will propose a more definite signal of the present 
ALP at Belle II, that is the polarization asymmetry in $\tau \to \mu(e) \gamma$, 
which can be a punchline of this tau-specific ALP with  the ``MFV".

\subsection{Polarization asymmetry in $\ell_{i} \rightarrow \ell_{j} \gamma$}

We first should observe that 
the present ALP couples to muon and electron, only through their right-handed chiral components, 
i.e., $\mu_R$ and $e_R$, because the couplings to light charged leptons 
arise only through the right-handed flavor rotation, as seen from 
Eqs.~(\ref{UR}), (\ref{mass}), and (\ref{gVA}). 
Thus, in the present framework of the ``MFV", 
the ALP couplings to light charged leptons significantly break 
the parity, hence would generate a sizable left-right asymmetry in the charged-lepton 
sector physics. 
Note that the chiral gauge interaction of the SM universally predicts predominantly left-handed polarization. 
In what follows, we shall show that the present ALP indeed predicts a sizable asymmetry in 
$\tau \to \mu (e) \gamma$, that can be detected as
the polarization asymmetry of the final state photon.

Quantification of the polarization asymmetry in $\tau \to \mu (e) \gamma$ 
can follow straightforwardly from that in $b \to s \gamma$, which has been 
discussed in the literature~\cite{Kou:2013gna,Haba:2015gwa}. 
Then, the Wilson coefficients (WCs) of dipole operators ($C_7$ and $C_7^{\prime}$) 
play 
the central role. 
For the photon polarization in $\ell_{i} \rightarrow \ell_{j} \gamma$ decay process, 
we define the relevant WCs, in a manner similar to those for $b \to s \gamma$~\cite{Kou:2013gna,Haba:2015gwa}, as follows: 
\begin{align}
\mathcal{L}_{\rm dipole}^{\ell} = \frac{G_F}{\sqrt{2}} (C_7^{\ell})_{ji} \frac{e}{16 \pi^2} m_{\ell_i} \left( \bar{\ell}_j \sigma^{\mu \nu} P_R \ell_i \right) F_{\mu \nu} + \frac{G_F}{\sqrt{2}} (C_7^{\prime \ell})_{ji} \frac{e}{16 \pi^2} m_{\ell_i} \left( \bar{\ell}_j \sigma^{\mu \nu} P_L \ell_i \right) F_{\mu \nu} \, ,
\label{eq:Ldipole}
\end{align}
where $\sigma^{\mu \nu} = \frac{i}{2} [ \gamma^{\mu}, \gamma^{\nu} ]$ 
and $P_{R/L} \equiv (1 \pm \gamma_5)/2$.  
Here we have ignored terms proportional to the lighter charged lepton mass, $m_{\ell_j} (\ll m_{\ell_i})$. 
Then the polarization parameter $\lambda_{\gamma}$ can be defined analogously to 
the $b \to s \gamma$ case~\cite{Kou:2013gna,Haba:2015gwa}, as 
\begin{align}
\lambda_{\gamma} 
= 
\frac{{\rm Re}\left[ (C_7^{\prime \ell})_{ji} / (C_7^{\ell})_{ji} \right]^2 + {\rm Im}\left[ (C_7^{\prime \ell})_{ji} / (C_7^{\ell})_{ji} \right]^2 - 1}{{\rm Re}\left[ (C_7^{\prime \ell})_{ji} / (C_7^{\ell})_{ji} \right]^2 + {\rm Im}\left[ (C_7^{\prime \ell})_{ji} / (C_7^{\ell})_{ji} \right]^2 + 1} \, .
\label{lambda-gamma}
\end{align}
 
The WCs in Eq.~(\ref{eq:Ldipole}) include contributions from both the SM 
and NP, which is the ALP in the present study. 
For a reference model, we consider the SM with massive Dirac neutrinos (denoted as SMD$\nu$). The model 
contribution for $\tau \rightarrow \mu \gamma$ can be estimated,  at the leading-nontrivial order in expansion with respect to $(m_\mu/m_\tau)$, as~\cite{Kou:2013gna} 
 \begin{align}
     (C_7^{\ell})_{\mu \tau}^{{\rm SM D} \nu}& \approx \frac{1}{2}A_{{\rm SM D} \nu}(x_{\nu}) \,, 
    \label{SM-C7}\\ 
     (C_7^{\prime \ell})_{\mu \tau}^{{\rm SM D} \nu}& \approx \frac{m_{\mu}}{m_{\tau}}(C_7^{\ell})_{\mu \tau}^{{\rm SM D} \nu}
\,, 
 \end{align}
where $x_{\nu}=m_{\nu}^2/m_{W}^2$, $m_{\nu}$ and $m_{W}$ are masses of an active neutrino and the $W$ boson,  respectively. Here we have simply taken the identical mass for all the active neutrinos, and the PMNS matrix to be unity, which, though being not precisely realistic, 
will not significantly affect the order of magnitude of estimate on the ${\rm SM D} \nu$ contribution. 
In Eq.~(\ref{SM-C7}) the loop function $A_{{\rm SM D} \nu(x)}$ reads 
\begin{align}
    A_{{\rm SM D} \nu}(x)=\frac{-8x^3-5x^2+7x}{12(x-1)^3}+\frac{3x^3-2x^2}{2(x-1)^4} \ln x
\,. 
\end{align}
Then we get the value of ${\rm SM D} \nu$ contribution on the WCs:
\begin{align}
     (C_7^{\ell})_{\mu \tau}^{{\rm SM D} \nu}& \approx -4.51\times10^{-27} \,, 
    \\ 
     (C_7^{\prime \ell})_{\mu \tau}^{{\rm SM D} \nu}& \approx -2.68 \times10^{-28}
\,. 
 \end{align}
Obviously, $| (C_7^{\ell})_{\mu \tau}^{{\rm SM D} \nu} | \gg  | (C_7^{\prime \ell})_{\mu \tau}^{{\rm SM D} \nu} |$ 
because $m_\mu/m_\tau \ll 1$, which is the consequence of the chirally left-handed gauged 
weak interaction, as noted above.

As to the NP contribution, 
the ALP arch and BZ graph amplitudes for $\ell_{i} \rightarrow \ell_{j} \gamma$ process are 
given in Refs.~\cite{Lindner:2016bgg} and~\cite{Cornella:2019uxs}, respectively. 
By translating those formulas back to the Lagrangian operator form, 
we find the following WCs:  
\begin{align}

(C_7^{\ell})_{ji}^{\rm NP} &=- \frac{Q^{\rm em}_{\ell}}{2f_a^2} \sum_f \Bigg[- (g_V^{\ell})_{jf} (g_V^{\ell})_{fi} I_{f, \, 1}^{++} + (g_A^{\ell})_{jf} (g_A^{\ell})_{fi} I_{f, \, 1}^{+-} - (g_A^{\ell})_{jf} (g_V^{\ell})_{fi} I_{f, \, 1}^{-+} + (g_V^{\ell})_{jf} (g_A^{\ell})_{fi} I_{f, \, 1}^{--}   \nonumber \\
&\left. - \frac{\alpha}{\pi} \frac{1}{(Q_{\ell_i}m_{\ell_i})^2 - (Q_{\ell_j}m_{\ell_j})^2} \left\{ (Q_{\ell_i}m_{\ell_i} - Q_{\ell_j}m_{\ell_j}) (g_A^{\ell})_{ji} + (Q_{\ell_i}m_{\ell_i} + Q_{\ell_j}m_{\ell_j}) (g_V^{\ell})_{ji} \right\} \frac{(g_A^f)_{f f}}{m_{f}}f\left(\frac{m_a^2}{m_{\ell_i}^2},\frac{m_a^2}{m_{f}^2}\right)  \right]  \, , \\[1.0ex]

(C_7^{\prime \ell})_{ji}^{\rm NP} &= - \frac{Q^{\rm em}_{\ell}}{2f_a^2} \sum_f  \Bigg[  - (g_V^{\ell})_{jf} (g_V^{\ell})_{fi} I_{f, \, 1}^{++} + (g_A^{\ell})_{jf} (g_A^{\ell})_{fi} I_{f, \, 1}^{+-} + (g_A^{\ell})_{jf} (g_V^{\ell})_{fi} I_{f, \, 1}^{-+} - (g_V^{\ell})_{jf} (g_A^{\ell})_{fi} I_{f, \, 1}^{--}   \nonumber \\
& \left. - \frac{\alpha}{\pi} \frac{1}{(Q_{\ell_i}m_{\ell_i})^2 - (Q_{\ell_j}m_{\ell_j})^2} \left\{ (Q_{\ell_i}m_{\ell_i} - Q_{\ell_j}m_{\ell_j}) (g_A^{\ell})_{ji} - (Q_{\ell_i}m_{\ell_i} + Q_{\ell_j}m_{\ell_j}) (g_V^{\ell})_{ji} \right\} \frac{(g_A^f)_{f f}}{m_{f}}f\left(\frac{m_a^2}{m_{\ell_i}^2},\frac{m_a^2}{m_{f}^2}\right)  \right]  \, , 
\end{align}
where the loop function $f(u,v)$ for the BZ diagram is given in Eq.(\ref{g2}).

\begin{figure}[htbp]
\centering
\subfigure[$C_7^{\prime \ell}$ arch]{
\begin{minipage}[t]{0.23\linewidth}
\centering
\includegraphics[width=1\linewidth]{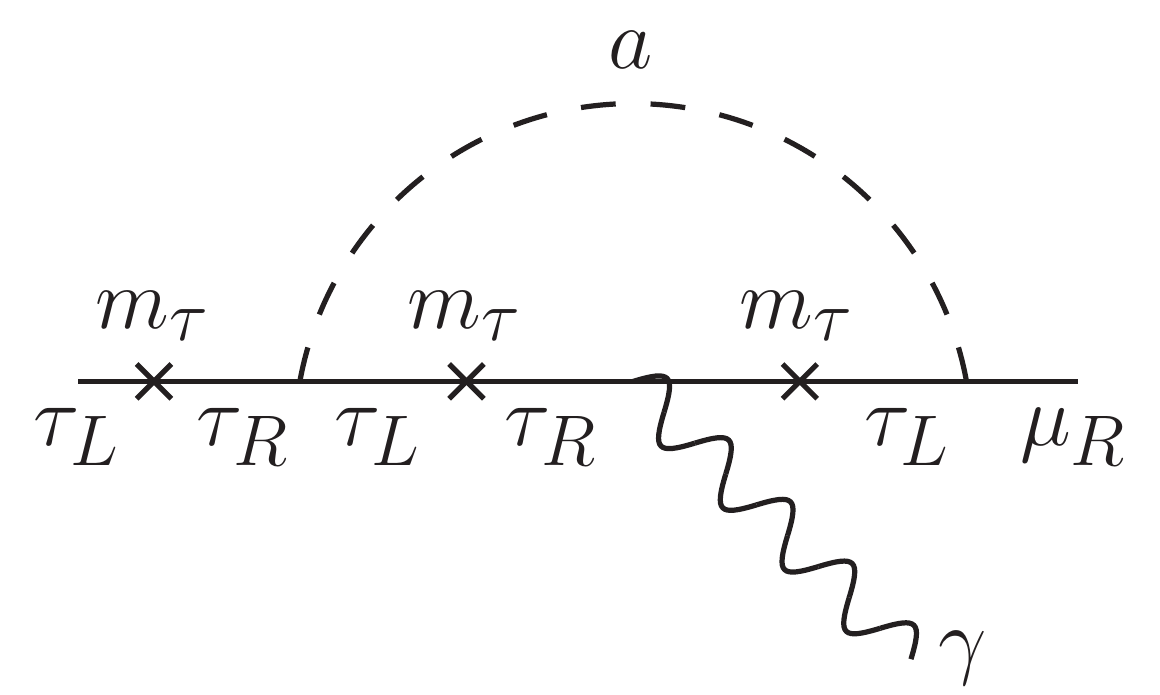}  
\end{minipage}%
}%
\subfigure[$C_7^{\prime \ell}$ BZ]{
\begin{minipage}[t]{0.23\linewidth}
\centering
\includegraphics[width=1\linewidth]{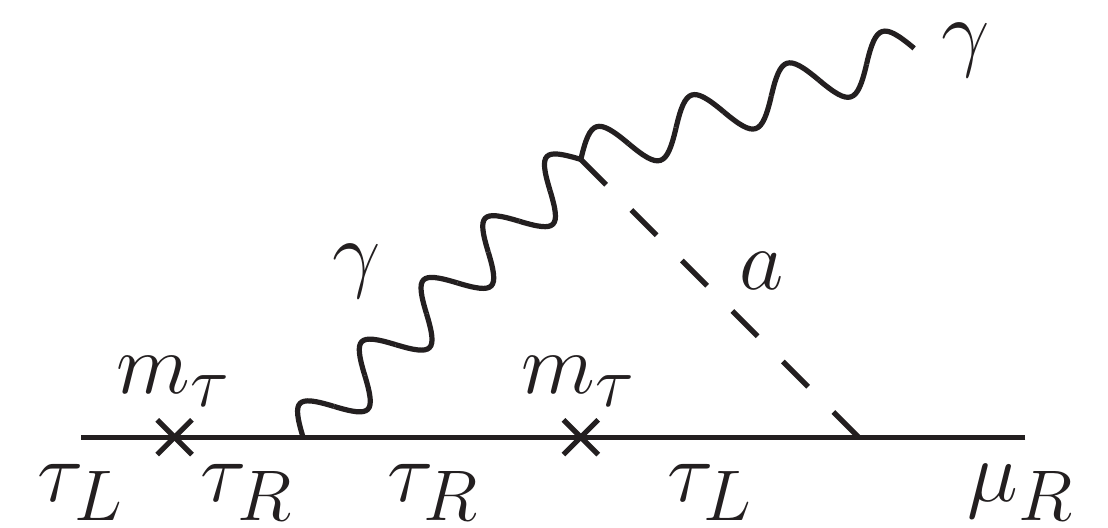}
\end{minipage}%
}%
\subfigure[$C_7^{\ell}$ arch: absent at the leading order]{
\begin{minipage}[t]{0.23\linewidth}
\centering
\includegraphics[width=1\linewidth]{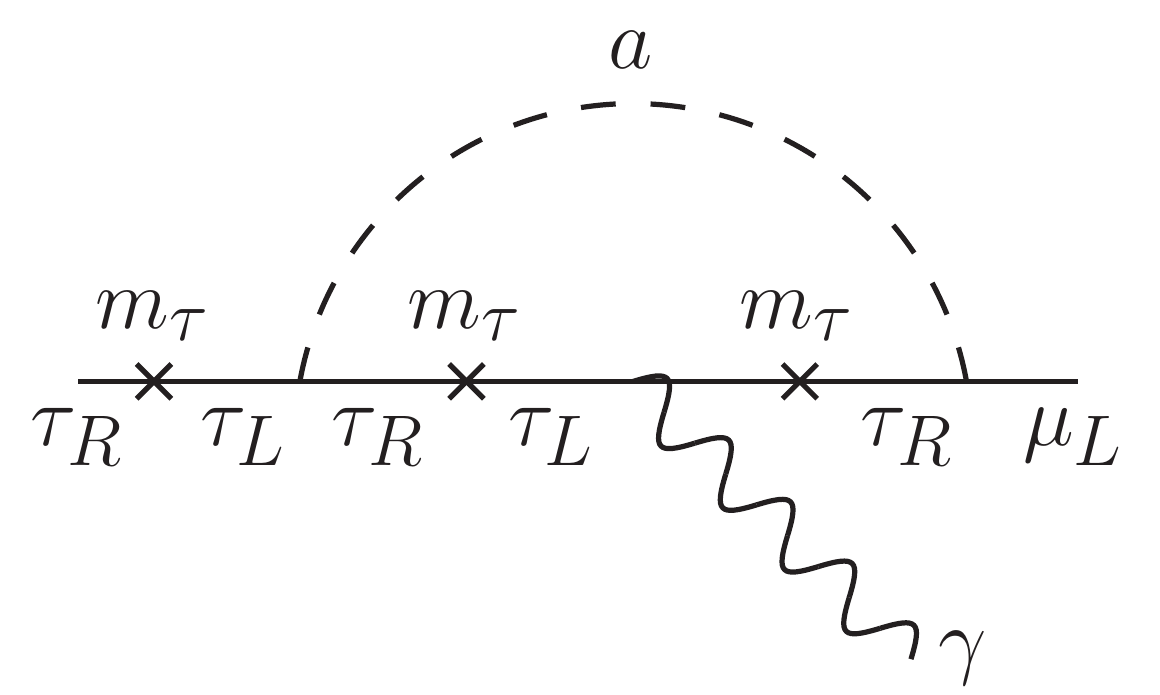}
\end{minipage}%
}%
\subfigure[$C_7^{\ell}$ BZ: absent at the leading order]{
\begin{minipage}[t]{0.23\linewidth}
\centering
\includegraphics[width=1\linewidth]{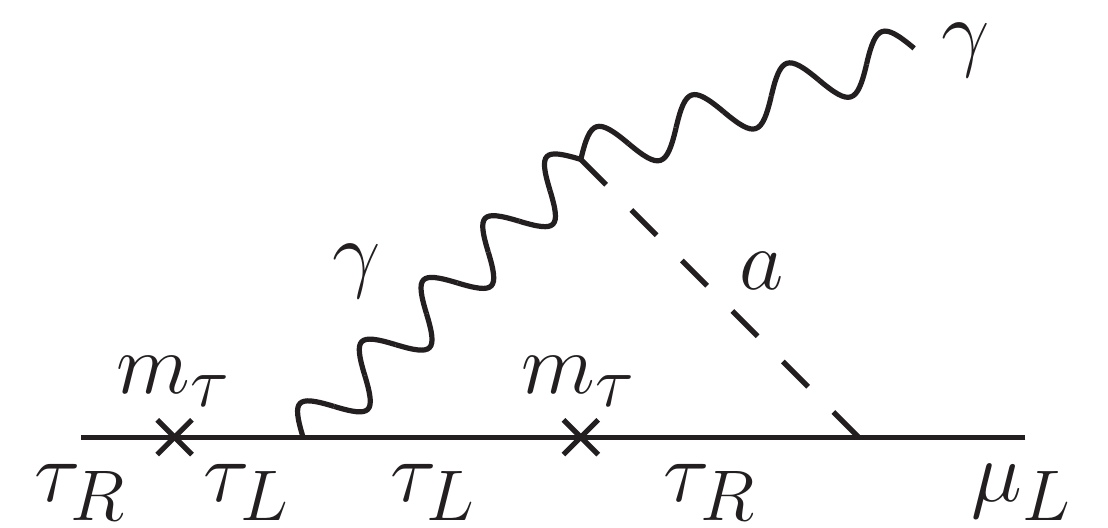}
\end{minipage}%
}%
\centering
\captionsetup{justification=raggedright}
\caption{The arch and BZ diagrams with chirality flips depicted 
contributing to $C_7$ and $C_7^{\prime}$ for $\tau \to \mu \gamma$, 
at the leading order in expansion with respect to  $(m_\mu/m_\tau)$. To this order, the diagrams (c) and (d) are not generated because 
of the ALP parity-breaking coupling property for muon. See the text, for more details.}
\label{C7}
\end{figure}

Since the present ALP is specifically coupled to $\mu_R$ and $e_R$, as emphasized above, 
we have $(C_7^{\ell})^{\rm NP} \approx 0$ at the nontrivial-leading order of $(m_\mu/m_\tau)$.  
Figure \ref{C7} would also help understand this point, where the case of $\tau \to \mu \gamma$ is exemplified. 
One can see that 
the nonzero contribution to $C_7$ necessarily requires the ALP coupling to $\mu_L$ (panels 
(c) and (d) in Fig.~\ref{C7}), which is forbidden at the leading order without 
suppression by extra chirality flip with $m_\mu$.  
On the other hand, the $C_7'$ does not yield the chirality flip of muon, 
as seen from the panel (a) and (b) in Fig.~\ref{C7}. 
A similar argument is applicable to the case of $\tau \to e \gamma$.


Thus, the present ALP with the ``MFV" and the maximal parity violation 
predicts 
\begin{align} 
 |(C_7^{\prime \ell})^{\rm NP} | \gg  |(C_7^{\ell})^{\rm NP}| 
\, , 
\end{align}
in sharp contrast to the ${\rm SM D} \nu$ prediction where $|(C_7^{\prime \ell})^{\rm SM} | \ll |(C_7^{\ell})^{\rm SM}|$, as noted above.  
We have found that the ALP contributions to the WCs 
are much greater than the SM contribution: $(C_7^{\prime \ell})^{\rm NP} = \mathcal{O}(10^{-13} -  10^{-12})$ in the surviving mass and $f_a$ ranges, see Table.~(\ref{ma-prediction}).  
Thereby, we find a punchline of the present ALP in the polarization asymmetry of 
Eq.~(\ref{lambda-gamma}): 
\begin{align} 
 \lambda_\gamma \approx 1 
\,, \label{lambda-gamma:ALP}
\end{align}
which shows the polarization trend completely opposite to the ${\rm SM D} \nu$'s one with $\lambda_\gamma^{\rm SM} \approx -1$. 


\subsection{Discrimination from other LFV NP candidates in light of Belle II} 

In this subsection, 
we compare phenomenological predictions from the present ALP to those 
from other LFV NP candidates with the same preferable mass scale as in Eq.~(\ref{ma-prediction}), 
within the prospected Belle II reach. 
Those NP include a light LFV $Z^{\prime}$ as in Ref.~\cite{Heeck:2016xkh}, 
and another type of flavorful ALP, 
as in Ref.~\cite{Cheung:2021mol}~%
\footnote{Another Belle II-target NP, which has the potential to explain two $(g-2)s$ for light charged leptons, 
would involve a light dark photon ($A^{\prime}$) 
with a mass range similar to the present ALP's, as discussed 
in Ref.~\cite{Mohlabeng:2019vrz} (300 MeV $\lesssim$ $m_{A^{\prime}}$ $\lesssim$ 1 GeV). 
This dark photon does not have LFV couplings, 
so the radiative tau-LFV signals as well as those polarization asymmetries ($\lambda_{\gamma})$ are the definite discriminators for the present ALP compared to the light dark photon. 
Furthermore, 
the light dark photon tends to favor about $2\sigma$ deviation for 
the (negatively pulled) old $\Delta \alpha_e|_{\rm old}$ in Eq.~(\ref{poae}), 
while the present ALP favors the positive value of $\Delta a_e$. 
Hence future experiments on determining the sign of $\Delta a_e$ can also distinguish the two particles.}.

\begin{itemize} 

\item 
In Ref.~\cite{Heeck:2016xkh}, 
the light $Z^{\prime}$ predominantly couples to muon and tau lepton. 
In their work, the preferred $Z^{\prime}$ mass range consistent with  the $2\sigma$ deviation range for $(g-2)_{\mu}$ has been found as $M_{Z^{\prime}}\simeq 2-900$ MeV,   
which is promising to be probed at Belle II. 
Our ``MFV" ALP would be a counter-candidate to this type of $Z^{\prime}$ in the high mass region in Belle II reach, both of which 
is accessible at the Belle II through similar LFV signals, as well as  
accounts for the current deviation in $(g-2)_{\mu}$.  
Of crucial is to notice, however, that our ALP can be clearly distinguished from $Z^{\prime}$ by the punchline, $\lambda_{\gamma}$, i.e., the polarization asymmetries of $\tau \rightarrow \mu \gamma$ and  $\tau \rightarrow e \gamma$ processes. 
As we discussed in the last section, 
the $L-R$ symmetry is significantly and maximally broken by the right-handed rotation 
due to the ``MFV" criterion, and highly polarized to right-handed, 
as evident from Eq.~(\ref{lambda-gamma:ALP}). 
In contrast, the LFV $Z^{\prime}$ is vertorlikely coupled to tau lepton and muon, 
so no $L-R$ asymmetry is created there. 
Thus the $\lambda_{\gamma}$ in $\tau\to \mu(e) \gamma$ 
is the definite discriminator of the present ``MFV" ALP from the light LFV $Z^{\prime}$.

\item 
Second, in a recent paper~\cite{Cheung:2021mol}, the authors focus on a light ALP 
having a high enough detection sensitivity at Belle II, through LFV and multi-lepton signals. Two parallel benchmark scenarios have been investigated in \cite{Cheung:2021mol},  where $c_{ee}$ and $c_{e \tau}$  or $c_{\mu \mu}$ and $c_{\mu \tau}$ are turned on, respectively. There are two possible discriminators to tell the present ``MFV" ALP from theirs. 
One is $\lambda_{\gamma}$, 
because the flavor couplings in their setup are totally independent with each other, 
not constrained by the ``MFV", so the parity violation is parametric, therefore  
no definite value of $\lambda_{\gamma}$ has been predicted in~\cite{Cheung:2021mol}. 
In contrast, the present ``MFV" ALP is definitely right-handed specific for muon and electron, 
where the parity is broken only by the right-handed rotation in the framework of the ``MFV", as 
emphasized in the previous section. 
The other discriminator is the tri-lepton signal where tau decays into three charged leptons. Since the ALP in~\cite{Cheung:2021mol} keeps the flavor-diagonal couplings to muon and electron, the tri-lepton process is possible to take place. In contrast, 
the present ALP is tau-specific, and has no flavor-diagonal coupling to muon and electron, 
as seen from Eq.~(\ref{couplings-hybrid}),  
so the tri-lepton signals are not produced. 

\end{itemize}

\section{Summary and Discussion}
\label{sec:summary}

In summary, we have discussed 
flavor-physics probes of 
an intrinsically flavorful ALP (
(``MFV"-ALP), with particular focus on 
a third-generation specific scenario, 
which generates the tau-lepton LFV processes, to be tested at the Belle II experiment.
The ALP is assumed to be tau-philic on the base where the PQ-like charge is defined, 
inspired by some folklore; ``{\it the third-generation is special}\,", 
and is allowed to also couple to muon and electron  
by the intrinsic ``MFV" arising from the right-handed flavor rotation within the SM.
Thus, the ALP couplings to muon and electron are right-handed specific.

We first employed two simplified and separated limits: 
the electron scenario (Sec.~\ref{E-scenario}) and muon scenario (Sec.~\ref{M-scenario}), 
as in Table~\ref{benTab}. 
We found that those scenarios are highly constrained by existing experimental 
limits from the LFV processes, in particular, $\tau \to e (\mu) \gamma$, 
and electron or muon $g-2$, so that the mixing angle $\theta_{13}$ 
or $\theta_{23}$ is required to extremely be close to $\pi/2$. 
This implies 
the preference of a mu or electron - tau flipped feature in 
the mass eigenbasis when coupled to the ALP. 

We then explored a hybrid scenario combining the two separated scenarios by turning 
on both $\theta_{13}$ and $\theta_{23}$. 
A fully viable parameter space on 
the ALP mass-photon coupling plane was found. See Fig.~\ref{parspace}. 
The ALP mass is limited in a range, $\sim (1.7 - 10)$ GeV 
and the ALP decay constant $f_a$ is constrained to be $\sim (12.8 - 67.9)$ GeV 
(Eq.~(\ref{ma-prediction})), if the same-sign multilepton signals constraint is disregarded. 
Remarkably, this ALP can be probed only by measurement 
of $\tau \rightarrow \mu \gamma$ and/or $\tau \rightarrow e \gamma$ at Belle II. %

We find that the same-sign multilepton signal at Belle II is a smoking-gun 
to probe the present ALP and
the polarization asymmetry in $\tau \rightarrow \mu \gamma$ and/or $\tau \rightarrow e \gamma$ is a punchline, 
which is definitely predicted to prefer 
the right-handed polarization due to the ``MFV" ($\lambda_\gamma \approx 1$ Eq.~(\ref{lambda-gamma:ALP})), 
overwhelming 
the prediction of the SM with massive Dirac neutrinos  
having the highly left-handed preference ($\lambda_\gamma^{\rm SM} \approx - 1$), in 
contrast to other light NP candidates which can also be promising to be probed at the Belle II, 
such as a light LFV $Z'$ and another flavorful ALP.

Several comments and discussions along the future prospect are in order.

\begin{itemize}

\item 

The coupling $g_{a\gamma\gamma}$ defined in Eq.~(\ref{g-gamma}) includes contribution from both bottom quark loops and tau lepton loops, because in the present study we have focused on a bottom-tau  specific ALP.  
However, even if other quarks are involved in the model by invoking other scenarios, 
the surviving parameter space (blue-shaded area in the center of Fig.~\ref{parspace}) will not substantially be changed, 
because the quark contributions are not constrained by all the leptonic processes. 
Other quark contributions will merely increase the value of $g_{a\gamma\gamma}$, 
without changing the allowed ALP mass and decay constant in Eq.~(\ref{ma-prediction}). 
Therefore, the blue shaded area in Fig.~\ref{parspace} will be just pushed upward, 
if other quarks come into the game. 
It would then be interesting to see how the surviving parameter space could get into 
the upper domain filled by the future Belle II prospect reach (at 50 ${\rm ab}^{-1}$), 
which is to be explored elsewhere. 

\item 
In cases of electron and muon scenarios, the ALP-quark coupling is necessary to keep 
$g_{a \gamma \gamma}$ to be sizable, 
since the ALP-lepton coupling almost vanishes to satisfy experimental constraints.~%
\footnote{Even in the hybrid case, the ALP-bottom quark coupling possibly takes nonzero without conflicting with the experimental constraints on $g_{a \gamma \gamma}$.}
Namely, the ALP-bottom quark coupling would lead to other probes for this third-generation specific scenario. 
For instance, $b \to d$ and $b \to s$ transitions can be induced dependent on the mixing angles,  
which would give us some interesting predictions on $B$ flavor physics. 
We, however, discard to discuss such possibilities, since it is beyond of current scope,  
and postpone elsewhere~%
\footnote{The flavor probes of ALPs with generic flavorful coupling are discussed in several articles~\cite{MartinCamalich:2020dfe, Calibbi:2020jvd,Bauer:2021mvw}.}.

\item Possible modeling to underlie the current third-generation specific ALP  

\begin{itemize}  

\item 

The present third-generation-philic ALP can be realized 
in a way similar to a class of variant QCD axion model with top-specific axion, 
based on a generic two Higgs doublet model, 
as discussed in Ref.~\cite{Chiang:2015cba}. 
In the reference the authors have also argued possible extension to 
give tau lepton the PQ charge, as well as top quark. 
As noted above, the bottom quark in the present ALP physics 
does not play a central role, and may even be replaced with top quark, 
as far as the lepton flavor physics is concerned. 
Therefore, the model setup in Ref.~\cite{Chiang:2015cba}
would straightforwardly lead to 
the present tau-specific ALP model. 
More precisely, to make the ALP mass much larger than a typical QCD axion mass scale, 
one needs to take the scale of 
a mass mixing term between two Higgs doublets, which explicitly breaks 
the PQ symmetry, to be 
somewhat large, and assume no CP violation in the Higgs sector,
and/or the second Higgs boson to be extremely heavy, 
 so that only the ALP is left in the light NP, without extra CP or parity violation.

\item 

The present ALP setup can also be naturally obtained when we consider the Froggatt-Nielsen (FN) mechanism~\cite{Froggatt:1978nt}~\footnote{The relationship 
between the PQ symmetry and a newly introduced global $U(1)$ symmetry 
for the FN mechanism has been addressed in Refs.~\cite{Davidson:1981zd,Wilczek:1982rv,Reiss:1982sq,Davidson:1983fy,Berezhiani:1990wn,Berezhiani:1990jj}. See e.g. Refs.~\cite{Ema:2016ops,Calibbi:2016hwq,Arias-Aragon:2017eww,Bjorkeroth:2017tsz,Linster:2018avp,Bjorkeroth:2018dzu,Bonnefoy:2019lsn,Carone:2020nlx,delaVega:2021ugs} for recent attempts.}. 
This mechanism can be used for realizing correct hierarchies of Yukawa couplings. 
In order to obtain appropriate SM fermion masses and CKM structure, Yukawa couplings ($Y_u$ for the up-quark sector; $Y_d$ for the down-quark sector; $Y_e$ for the charged lepton sector) should roughly have the following texture form:
\begin{align}
Y_u \sim \begin{pmatrix}
\lambda^6 & \lambda^5 & \lambda^3 \\
\lambda^5 & \lambda^4 & \lambda^2 \\
\lambda^3 & \lambda^2 & 1
\end{pmatrix}, \quad Y_d \sim \begin{pmatrix}
\lambda^6 & \lambda^{5.5} & \lambda^5 \\
\lambda^5 & \lambda^{4.5} & \lambda^4 \\
\lambda^3 & \lambda^{2.5} & \lambda^2
\end{pmatrix}, \quad Y_e \sim \begin{pmatrix}
\lambda^6 & \lambda^5 & \lambda^3 \\
\lambda^{5.5} & \lambda^{4.5} & \lambda^{2.5} \\
\lambda^5 & \lambda^4 & \lambda^2
\end{pmatrix},
\label{eq:roughYukawa}
\end{align}
where $\lambda \simeq 0.22$. 
This can be easily obtained when we assume an anomalous $U(1)$ charge assigned for SM particles~\footnote{
The gauge anomaly associated with this new $U(1)$ gauge symmetry can be canceled by the Green-Schwarz mechanism~\cite{Green:1984sg}. 
} as follows: 
\begin{align}
Q_i &: (3, 2, 0), \quad u_{R \, i}^c : (3, 2, 0), \quad d_{R \, i}^c : (3, 2.5, 2), \nonumber \\
L_i &: (3, 2.5, 2), \quad e_{R \, i}^c : (3, 2, 0), \quad H : 0
\,, 
\end{align}
where $Q_i$ and $L_i$ are quark and lepton doublets, respectively; $i$ denotes 
the generation index; the upper script $c$ attached on fields stands for the charge conjugation; $H$ denotes the Higgs doublet.

In order to write down the Yukawa couplings, we need to introduce a new scalar field, $\Theta$, which is singlet under SM gauge symmetries and has the anomalous $U(1)$ charge $-1$. 
Once $\Theta$ acquires the vacuum expectation value like $\langle \Theta \rangle = \lambda \Lambda$, where $\Lambda$ is some cutoff scale, one can reproduce the hierarchical Yukawa structure in Eq.~\eqref{eq:roughYukawa}.

In this framework, the ALP-fermion couplings follow the same hierarchies 
as in Eq.~\eqref{eq:roughYukawa}, which, therefore, automatically leads to 
the third-generation-specific ALP.  
When we assume the ALP to have the anomalous $U(1)$ charge $-2$, only ALP couplings to bottom and tau lepton arise as $O(1)$, and the other couplings are smaller than $\lambda^2 \sim 0.05$~\footnote{In the case of non-supersymmetric models, the ALP-top coupling in this framework is obtained from 
\begin{align}
\left( \frac{\Theta^{\dagger}}{\Lambda} \right)^2 m_t \, a \, \bigl( \bar{t} \gamma_5 t \bigr) \rightarrow \lambda^2 \, m_t \, a \, \bigl( \bar{t} \gamma_5 t \bigr).
\end{align}
In contrast, such terms are absent in supersymmetric models since we cannot use $\Theta^{\dagger}$ in the superpotential. }. 
Thus, the presently analyzed ALP coupling properties can be 
thought of as an extreme limit of those arising from this FN modeling. 
We have indeed checked that if we turn on the other ALP-fermion couplings with proper sizes as predicted from the FN mechanism, 
the results present in the main text will not substantially  
be changed~\footnote{Actually, nonzero couplings of $Q_e$ and $Q_{\mu}$ induce a $c_{e \mu}$ coupling in the case of the hybrid scenario, which is severely constrained by LFV processes related to the muon. However, these constraints can be easily avoided by setting $\theta_{13} \sim 0$. Even in this case, the main results of the present paper will not essentially be changed.}.

\end{itemize}


\end{itemize}

\section*{Acknowledgements}

We thank Motoi Endo, Syuhei Iguro, and Robert Ziegler for useful comments and discussions. 
S.M. work was supported in part by the National Science Foundation of China (NSFC) under Grant No.11747308, 11975108, 12047569 and the Seeds Funding of Jilin University. 
The work of C.C. has been partially supported by the TAQ honors program in science from the Office of Undergraduate Education and College of Physics in Jilin University.  

\end{document}